\documentclass[aps,prd,preprintnumbers,showpacs,showkeys,nofootinbib,
superscriptaddress,fleqn,floatfix,tightenlines,10pt]{revtex4-1}
\usepackage{amsmath,amsfonts,amssymb,amscd,amsxtra,amsthm}
\usepackage{graphicx}  
\usepackage{epstopdf}
\usepackage{dcolumn}  
\usepackage{bm}          
\usepackage{slashed}
\usepackage{cancel}
\usepackage{float} 
\usepackage{mathtools}
\usepackage{amsbsy}
\usepackage{amstext}

\usepackage[utf8]{inputenc} 
\usepackage{booktabs}
\usepackage[normalem]{ulem} 
\usepackage[dvipsnames]{xcolor} 
\usepackage{tabularx}
\usepackage{enumitem}  
\usepackage{array} 
\usepackage{slashed}
\usepackage{tikz}
\usepackage{float}
\usepackage{multirow}
\renewcommand\sout{\bgroup \color{red} \ULdepth=-.5ex \ULset}

\makeatletter

\begin{document}  
\title{Gravitational form factors of a baryon with spin-3/2 }
\author{June-Young Kim}
\email[E-mail: ]{Jun-Young.Kim@ruhr-uni-bochum.de}
\affiliation{Ruhr-Universit\"at Bochum, Fakult\"at f\"ur Physik und Astronomie,
Institut f\"ur Theoretische Physik II, D-44780 Bochum, Germany}
\author{Bao-Dong Sun}
\email[E-mail: ]{sunbd@sdu.edu.cn}
\affiliation{Key Laboratory of Particle Physics and Particle Irradiation (MOE),
Institute of Frontier and Interdisciplinary Science,
Shandong University, Qingdao, Shandong 266237, China}
\date{\today}
\begin{abstract}
The energy-momentum tensor (EMT) for a spin-3/2 baryon is related to  \emph{seven} mechanical quantities. In this work, we provide the general form of the gravitational form factors (GFFs) for a spin-3/2 baryon by using the multipole expansion and interesting relations between the EMT densities and the GFFs. To verify those general relations, we study the nucleon and the $\Delta$ GFFs within the SU(2) Skyrme model based on the large $N_{c}$ limit.
\end{abstract}
\maketitle
\section{Introduction \label{sec:1}}
The energy-momentum tensor (EMT) of the nucleon contains fundamental  information on its three mechanical properties, i.e., the mass, spin, and $D$-term of the nucleon. These mechanical properties are related to three gravitational form factors (GFFs) at zero momentum transfer. While the mass and spin of the nucleon  are relatively well known, the $D$-term, which provides essential information on how the nucleon experiences the internal force by its constituents, is much less known~\cite{Polyakov:2002yz,  Polyakov:2018zvc, Lorce:2018egm}. Thus, it is of great importance to  investigate the $D$-term of the  nucleon.  The GFFs were at first  considered~\cite{Kobzarev:1962wt, Pagels:1966zza} as a merely academic subject, since it was impossible to measure them directly. The reason is that the  graviton indeed weakly interacts with the nucleon. However, the generalized parton distributions (GPDs) shed novel light on the GFFs of the nucleon, since the GFFs are  identified as the second Mellin moments of the unpolarized GPDs~\cite{Ji:1996ek,
  Polyakov:1999gs, Diehl:2003ny, Goeke:2001tz}, which get accessible experimentally in hard exclusive reactions. In fact, there are several ongoing and planned facilities that are ideal to measure the GPDs,
such as the newly upgraded 12 GeV Continuous Electron Beam Accelerator Facility (CEBAF) at the Jefferson Lab, the Electron-Ion Collider (EIC) that will be constructed at Brookhaven National Laboratory (BNL)~\cite{Proceedings:2020eah}, and the
planned Electron-Ion Collider in China (EicC)~\cite{Chen:2018wyz}. The EIC experiments will produce unprecedented experimental results on the GPDs~\cite{Accardi:2012qut}. 

Another interesting object is the $\Delta$ isobar, which is the first excited baryon
with spin-3/2. While the electromagnetic properties of the $\Delta$ have been
extensively investigated both experimentally and theoretically, there are only a
few theoretical works on the GFFs of the $\Delta$. 
In Ref.~\cite{Cotogno:2019vjb}, the relevant structure for the GFFs of the
$\Delta$ was sorted out. The general properties of the pressure and shear forces
for the $\Delta$ isobar were derived in the large-$N_{c}$ limit~\cite{Panteleeva:2020ejw}. 
Compared to the GFFs of the nucleon, the GFFs of the $\Delta$ isobar have a far
more complicated structure: \emph{ten} different kinds of the GFFs together with the
EMT-nonconserving form factors. Thus, it is rather difficult to grasp the physical meaning of the GFFs of the $\Delta$ isobar. In this sense, the multipole expansion of the EMT matrix elements of the $\Delta$ will reveal the physical implications of the $\Delta$ GFFs. The GFFs of the nucleon were first extracted from the
GPDs~\cite{Burkert:2018bqq}, by using the data on deeply virtual Compton scattering
(DVCS) of the nucleon. On the other hand, it is rather difficult to measure experimentally the GFFs of the $\Delta$ or to extract them from the corresponding GPDs because of its short-lived nature. 
In the meantime, we anticipate that lattice QCD will provide a clue to understanding the
GFFs of the $\Delta$.   

In the present work, we aim at how the EMT matrix elements of the $\Delta$ can be
compactly expressed in terms of the GFFs, using the multipole expansion. 
By doing that, we are able to find interesting relations between the EMT densities
and these GFFs. We want to emphasize that the relations obtained in the present work
are model-independent. We will verify these general relations within the framework of
the SU(2) Skyrme model. The model is known to be one of the simplest ones
for describing the lowest-lying baryons based on the large $N_c$ expansion.
In the limit of $N_c\to\infty$~\cite{Witten:1979kh,  Witten:1983tx}, a baryon arises as
a topological soliton with an effective mesonic degrees of freedom.
In addition, the model satisfies the essential properties of QCD such as chiral symmetry and
its spontaneous breaking. The model describes very well important properties of the
nucleon in low-energy regimes. Furthermore, the model has explained numerous
observables for the baryons~\cite{Holzwarth:1985rb, Zahed:1986qz} and has even described well 
general properties of the nucleon GFFs. Thus, we will use the Skyrme model to examine
those of the $\Delta$ GFFs.

There are various works on the GFFs for a hadron with different spins. The
parametrizations of the GFFs for a hadron with various spin are discussed in
Ref.~\cite{Pagels:1966zza, Holstein:2006ud, Polyakov:2019lbq, Cosyn:2019aio,
  Cotogno:2019vjb}. The GFFs of the spin-0 hadrons are investigated in the chiral quark model~\cite{Broniowski:2008hx, Son:2014sna}, lattice QCD~\cite{Shanahan:2018pib},  the parameter fitting from experimental
data~\cite{Kumano:2017lhr}, and  the Nambu–Jona-Lasinio (NJL)
model~\cite{Freese:2019bhb}. For the spin-1/2 hadrons, there are results from the
effective chiral theory~\cite{Alharazin:2020yjv}, the chiral quark-soliton
model~\cite{Goeke:2007fp, Kim:2020nug}, the SU(2) Skyrme
model~\cite{Cebulla:2007ei, Kim:2012ts}, the $\pi\mathrm{-}\rho\mathrm{-}\omega$
model~\cite{Jung:2013bya, Jung:2014jja}, the bag model~\cite{Neubelt:2019sou},
the QCD sum rule{~\cite{Anikin:2019kwi, Azizi:2019ytx} and the
  lattice   QCD~\cite{Shanahan:2018nnv, Detmold:2019ghl}.
  The renormalization group properties of the nucleon’s twist-four GFF
  $\bar{c}_{q,g}(t)$~\footnote{ $\bar{c}_{q,g}(t)=2M_N F^T_{3,0}(t)$ in the notation of Ref.~\cite{Cotogno:2019vjb} where $M_N$ is the nucleon mass. },
which plays a role in the nucleon’s transverse spin sum rule, are studied in Ref.~\cite{Hatta:2018sqd}.
In the case of the spin-1 hadrons, the relations between the GPDs and the GFFs
are established in Ref.~\cite{Cosyn:2018thq, Cosyn:2019aio, Polyakov:2019lbq}
and the GFFs of the vector mesons are evaluated in
Ref.~\cite{Abidin:2008ku,Freese:2019bhb, Sun:2020wfo}.
For the higher spin hadrons, the works on the parametrization of the stress
tensor are made in Ref.~\cite{Polyakov:2018rew, Panteleeva:2020ejw}. 

We sketch the present work as follows: In Section \ref{sec:2} we define
the hadronic matrix elements of the EMT as the GFFs and reorganize the GFFs
in terms of the multipole expansion. We also define the EMT densities of a baryon
with spin-3/2 in terms of the multipole expansion and present the relations
between the GFFs and the EMT densities. In order to verify the general requirements
and relations proposed in Section \ref{sec:2}, the GFFs and the EMT multipole
densities of the $\Delta$ are obtained within the Skyrme model in
Section \ref{sec:3} and the numerical results are showed and discussed in
Section \ref{sec:4}. In the final Section~\ref{sec:5}, we present a summary and conclusions. 

\section{Gravitational form factors of a spin-3/2 hadron \label{sec:2}}
We use the covariant normalization $\langle p', \sigma'| p, \sigma \rangle = 2p^{0} (2\pi)^{3}\delta_{\sigma' \sigma} \delta^{(3)}(\bm{p}'-\bm{p})$ of one-particle states, and introduce kinematical variables $P^{\mu}=(p^{\mu}+p'^{\mu})/2$, $\Delta^{\mu}=p'^{\mu}-p^{\mu}$ and $\Delta^{2} = t$.
For the GFFs of a spin-3/2 particle in QCD, the matrix elements of EMT current is given by~\cite{Cotogno:2019vjb}\footnote{In order to be in line with the definition of the matrix elements of the EMT current for a spin-1/2 baryon, we reparametrized the expressions given in Ref.~\cite{Cotogno:2019vjb} as $(F^{a}_{1,0},F^{a}_{1,1},F^{a}_{2,0},F^{a}_{2,1},F^{a}_{3,0},F^{a}_{3,1},F^{a}_{4,0},F^{a}_{4,1},F^{a}_{5,0},F^{a}_{6,0})=(F^{T}_{1,0},F^{T}_{1,1},4F^{T}_{2,0},4F^{T}_{2,1},F^{T}_{3,0},F^{T}_{3,1},\frac{1}{2}F^{T}_{4,0},\frac{1}{2}F^{T}_{4,1},\frac{1}{2}F^{T}_{5,0},\frac{1}{2}F^{T}_{6,0})$. Note that there is a typo in Ref.~\cite{Cotogno:2019vjb}, and it should be corrected as $g^{\mu\nu}\Delta_{\alpha'}\Delta_{\alpha} F^{T}_{5,0}\to 2g^{\mu\nu}\Delta_{\alpha'}\Delta_{\alpha}F^{T}_{5,0}$.}
\begin{align}
\langle p' ,\sigma' | \hat{T}^{\mu\nu}_{a}(0) | p , \sigma  \rangle = -\overline{u}^{\alpha'}(p',\sigma') &\bigg{[}\frac{ P^{\mu} P^{\nu}}{m} \left(g_{\alpha'\alpha}F^{a}_{1,0}(t) - \frac{\Delta_{\alpha'} \Delta_{\alpha}}{2m^{2}}F^{a}_{1,1}(t)\right) \cr
&+\frac{(\Delta^{\mu}\Delta^{\nu}-g^{\mu\nu}\Delta^{2})}{4m}\left(g_{\alpha'\alpha}F^{a}_{2,0}(t) - \frac{\Delta_{\alpha'} \Delta_{\alpha}}{2m^{2}}F^{a}_{2,1}(t)\right) \cr
&+mg^{\mu\nu}\left(g_{\alpha'\alpha}F^{a}_{3,0}(t) - \frac{\Delta_{\alpha'} \Delta_{\alpha}}{2m^{2}}F^{a}_{3,1}(t)\right) \cr
&+\frac{i}{2}\frac{(P^{\mu}\sigma^{\nu\rho}+P^{\nu}\sigma^{\mu\rho})\Delta_{\rho}}{m}\left(g_{\alpha'\alpha}F^{a}_{4,0}(t) - \frac{\Delta_{\alpha'} \Delta_{\alpha}}{2m^{2}}F^{a}_{4,1}(t)\right) \cr
&-\frac{1}{m}(\Delta^{\mu}g^{\nu}_{\alpha'} \Delta_{\alpha} + \Delta^{\nu}g^{\mu}_{\alpha'} \Delta_{\alpha} + \Delta^{\mu}g^{\nu}_{\alpha} \Delta_{\alpha'} + \Delta^{\nu}g^{\mu}_{\alpha} \Delta_{\alpha'} \cr
&-2g^{\mu\nu}\Delta_{\alpha'}\Delta_{\alpha} - g^{\mu}_{\alpha'}g^{\nu}_{\alpha}\Delta^{2} - g^{\nu}_{\alpha'}g^{\mu}_{\alpha}\Delta^{2}) F^{a}_{5,0}(t) \cr
&+m  (g^{\mu}_{\alpha'}g^{\nu}_{\alpha} + g^{\nu}_{\alpha'}g^{\mu}_{\alpha})F^{a}_{6,0}(t)\bigg{]}u^{\alpha}(p,\sigma) 
\end{align}
where ${u}^{\alpha}(p,\sigma)$ is the Rarita-Schwinger spinor, and it satisfies the Dirac equation $(\slashed{p}-m)u^{\alpha}(p,\sigma)=0$ and the subsidiary conditions $\gamma_{\alpha} u^{\alpha}(p,\sigma)=0$ and $p_{\alpha} u^{\alpha}(p,\sigma)=0$. Here, $\sigma (\sigma')$ is the initial (final) spin projections. The normalization of the Rarita-Schwinger spinors is taken to be $\overline{u}^{\alpha'}_{\sigma'}(p)g_{\alpha'\alpha}u^{\alpha}_{\sigma}(p) = -2m \delta_{\sigma'\sigma}$. The index $a$ runs from a gluon to quark flavors. The quark and gluon form factors $F^{a}_{i,k}(i=1,2,4,5)$ are individually conserved, whereas $F^{a}_{i,k}(i=3,6)$ are not conserved. Note that we name the GFFs of a baryon with spin-3/2 according to Ref.~\cite{Cotogno:2019vjb} and reparametrize them to be analogous with those with spin-1/2 and spin-1. The separate quark and gluon GFFs depend on the renormalization scale $\mu$, which is suppressed for simplicity. Because of the EMT conservation, the non-conservation terms $F^{a}_{i,k}(i=3,6)$ have constraints, i.e., $\sum_{a}F^{a}_{i,k}=0  \, (i=3,6)$. The scale-invariant total GFFs are obtained as $F_{i,k}=\sum_{a} F^{a}_{i,k} \, (i=1,2,4,5)$.

\subsection{Gravitational form factors in Breit frame}
Before discussing the GFFs, we define the $n$-rank irreducible tensors and the multipole operators. The $n$-rank irreducible tensors in coordinate (or momentum) space are given by
\begin{align}
Y^{i_{1}i_{2}...i_{n}}_{n}(\Omega_{r}) = \frac{(-1)^{n}}{(2n-1)!!}r^{n+1}\partial^{i_{1}}\partial^{i_{2}}...\partial^{i_{n}}\frac{1}{r}, \ \ \ \ \ Y^{i_{1}i_{2}...i_{n}}_{n}(\Omega_{p}) = \frac{(-1)^{n}}{(2n-1)!!}p^{n+1}\partial^{i_{1}}\partial^{i_{2}}...\partial^{i_{n}}\frac{1}{p}.
\end{align}
Thus, one gets the following expressions as
\begin{align}
Y_{0}(\Omega_{r})=1, \ \  Y^{i}_{1}(\Omega_{r})=\frac{r^{i}}{r},  \ \ Y^{ij}_{2}(\Omega_{r})=\frac{r^{i}r^{j}}{r^{2}}-\frac{1}{3}\delta^{ij}, \ \ Y^{ijk}_{3}(\Omega_{r})=\frac{r^{i}r^{j}r^{k}}{r^{3}}-\frac{1}{5}\left(\frac{r^{k}}{r}\delta^{ij} + \frac{r^{i}}{r}\delta^{jk} + \frac{r^{j}}{r}\delta^{ik} \right).
\label{eq:tensor}
\end{align}
For a hadron with spin-3/2, the quadrupole- and octupole-spin operators $\hat{Q}^{ij}$(rank 2 tensor) and $\hat{O}^{ijk}$(rank 3 tensor) are respectively defined in terms of the spin operator $\hat{S}^{i}$ as
\begin{align}
\hat{Q}^{ij} &= \frac{1}{2}\left( \hat{S}^{i}\hat{S}^{j} +\hat{S}^{j}\hat{S}^{i} -\frac{2}{3}S(S+1)\delta^{ij}\right), \cr
\hat{O}^{ijk} &= \frac{1}{6}\bigg{(} \hat{S}^{i}\hat{S}^{j}\hat{S}^{k}+\hat{S}^{j}\hat{S}^{i}\hat{S}^{k}+\hat{S}^{k}\hat{S}^{j}\hat{S}^{i}+\hat{S}^{j}\hat{S}^{k}\hat{S}^{i}+\hat{S}^{i}\hat{S}^{k}\hat{S}^{j}+\hat{S}^{k}\hat{S}^{i}\hat{S}^{j}  \cr
&\hspace{0.1cm}-\frac{6S(S+1)-2}{5}(\delta^{ij}\hat{S}^{k}+\delta^{ik}\hat{S}^{j}+\delta^{kj}\hat{S}^{i})\bigg{)},
\end{align}
with $i,j,k=1,2,3$, and the operators are symmetrized and traceless ($\hat{Q}^{ii}=0 \mathrm{ \ and}  \ \hat{O}^{iij}=\hat{O}^{iji}=\hat{O}^{jii}=0$). The spin operators can be expressed in terms of the SU(2) Clebsch-Gordan coefficients in the spherical basis as
\begin{align}
\hat{S}^{a}_{\sigma'\sigma} = \sqrt{S(S+1)}C^{S \sigma'}_{S \sigma 1 a} \ \ \ \mathrm{with} \ \ \ (a=0,\pm1. \  \  \sigma,\sigma'=0, \cdot\cdot\cdot,\pm S).
\end{align}
In the Breit frame the average of the baryon momenta and the momentum transfer are respectively defined by $P^{\mu}=(p^{\mu}+p'^{\mu})/2=(E,0,0,0)$ and $\Delta^{\mu}= p'^{\mu}-p^{\mu}=(0,\bm{\Delta})$ with the initial (final) momentum $p\,(p')$. The momentum squared is defined as $\Delta^{2}=-\bm{\Delta}^{2} = t = 4(m^{2}-E^{2})$ with the baryon mass $m$. The explicit expressions of the Rarita-Schwinger spinor and the polarization vetor are given in Appendix~\ref{Appendix:A}. In this frame, the matrix elements of the EMT current are expressed in terms of the gravitational multipole form factors (GMFFs) as
\begin{align}
\langle p', \sigma' | \hat{T}^{00}_{a}(0) | p , \sigma \rangle &= 2mE \bigg{[} \mathcal{E}^{a}_{0}(t)\delta_{\sigma'\sigma} + \left(\frac{\sqrt{-t}}{m}\right)^{2} \hat{Q}_{\sigma'\sigma}^{kl}Y^{kl}_{2} \mathcal{E}^{a}_{2}(t)\bigg{]}, \cr
\langle p', \sigma' | \hat{T}^{0i}_{a}(0) | p , \sigma \rangle &= 2mE\bigg{[} \frac{\sqrt{-t}}{m}i\epsilon^{ikl}Y_{1}^{l}\hat{S}^{k}_{\sigma'\sigma}\mathcal{J}^{a}_{1}(t) +\left(\frac{\sqrt{-t}}{m}\right)^{3}i\epsilon^{ikl}Y^{lmn}_{3}\hat{O}^{kmn}_{\sigma'\sigma}\mathcal{J}^{a}_{3}(t)\bigg{]}, \cr
\langle p', \sigma' | \hat{T}^{ij}_{a}(0) | p , \sigma \rangle &=2mE\bigg{[} \frac{1}{4m^{2}}(\Delta^{i}\Delta^{j} + \delta^{ij}\Delta^{2}) D^{a}_{0}(t)\delta_{\sigma'\sigma }+ \frac{1}{4m^{4}} \hat{Q}^{kl}_{\sigma'\sigma}(\Delta^{i}\Delta^{j} + \delta^{ij}\Delta^{2}) \Delta^{k} \Delta^{l}D^{a}_{3}(t) \cr
&\hspace{1cm}+ \frac{1}{2m^{2}} \left(\hat{Q}^{ik}_{\sigma'\sigma} \Delta^{j}\Delta^{k} +\hat{Q}^{jk}_{\sigma'\sigma}  \Delta^{i}\Delta^{k} +  \hat{Q}^{ij}_{\sigma'\sigma} \Delta^{2} - \delta^{ij}\hat{Q}^{kl}_{\sigma'\sigma} \Delta^{k}\Delta^{l} \right)D^{a}_{2}(t) \cr
&\hspace{1cm}+\delta_{\sigma'\sigma}\delta^{ij}\left(-F^{a}_{3,0}(t)-\frac{2}{3}F^{a}_{6,0}(t) +\frac{t}{6m^{2}}F^{a}_{3,0}(t) + \frac{t}{6m^{2}}
F^{a}_{3,1}(t) - \frac{t^{2}}{24m^{4}}F^{a}_{3,1}(t)\right) \cr
&\hspace{1cm}+ \delta_{\sigma'\sigma}\Delta^{i}\Delta^{j}\left(-\frac{F^{a}_{6,0}(t)}{6m^{2}}\right)+\left(\hat{Q}^{ik}_{\sigma'\sigma} \Delta^{j}\Delta^{k} + \hat{Q}^{jk}_{\sigma'\sigma} \Delta^{i}\Delta^{k} \right) \frac{F^{a}_{6,0}(t)}{6m(m+E)} + \frac{2}{3}\hat{Q}^{ij}_{\sigma'\sigma}F^{a}_{6,0}(t) \cr
&\hspace{1cm}+ \delta^{ij}\hat{Q}^{kl}_{\sigma'\sigma} \Delta^{k}\Delta^{l} \left( \frac{F^{a}_{3,0}(t)}{6m^{2}} +\frac{F^{a}_{3,1}(t)}{6m^{2}} - \frac{F^{a}_{3,1}(t) t}{24m^{4}}\right) +\hat{Q}^{kl}_{\sigma'\sigma} \Delta^{k}\Delta^{l}\Delta^{i}\Delta^{j} \frac{F^{a}_{6,0}(t)}{24m^{2}(m+E)^{2}}    \bigg{]}, 
 \label{eq:1}
\end{align}
with
\begin{align}
\mathcal{E}^{a}_{0}(t) &= F^{a}_{1,0}(t) + F^{a}_{3,0}(t) \cr
&+\frac{t}{6m^{2}}\bigg{[}-\frac{5}{2}F^{a}_{1,0}(t) - F^{a}_{1,1}(t) -\frac{3}{2} F^{a}_{2,0}(t) +4 F^{a}_{5,0}(t) + 3F^{a}_{4,0}(t)  -F^{a}_{3,0}(t) - F^{a}_{3,1}(t) - F^{a}_{6,0}(t) \bigg{]}\cr
&+\frac{t^{2}}{12m^{4}}\bigg{[}\frac{1}{2}F^{a}_{1,0}(t) +   F^{a}_{1,1}(t) +\frac{1}{2} F^{a}_{2,0}(t) +\frac{1}{2} F^{a}_{2,1}(t) -4 F^{a}_{5,0}(t) -F^{a}_{4,0}(t) -F^{a}_{4,1}(t)  +\frac{1}{2}F^{a}_{3,1}(t)  \bigg{]} \cr
&+\frac{t^{3}}{48m^{6}}\bigg{[}- \frac{1}{2}F^{a}_{1,1}(t) -\frac{1}{2} F^{a}_{2,1}(t)  +F^{a}_{4,1}(t) \bigg{]}, \cr
\mathcal{E}^{a}_{2}(t) &= -\frac{1}{6} \bigg{[} F^{a}_{1,0}(t) + F^{a}_{1,1}(t)  -4 F^{a}_{5,0}(t) + F^{a}_{3,0}(t) + F^{a}_{3,1}(t) + F^{a}_{6,0}\bigg{]} \cr 
& +\frac{t}{12m^{2}}\bigg{[}\frac{1}{2}F^{a}_{1,0}(t)+F^{a}_{1,1}(t) + \frac{1}{2} F^{a}_{2,0}(t) + \frac{1}{2} F^{a}_{2,1}(t)  -4 F^{a}_{5,0}(t) - F^{a}_{4,0}(t) -F^{a}_{4,1}(t) +\frac{1}{2}F^{a}_{3,1}(t)  \bigg{]} \cr
& +\frac{t^{2}}{48m^{4}} \bigg{[} -\frac{1}{2}F^{a}_{1,1}(t) -\frac{1}{2} F^{a}_{2,1}(t) +F^{a}_{4,1}(t)\bigg{]},
\end{align}
\begin{align}
\mathcal{J}^{a}_{1}(t) &=\frac{1}{3} \bigg{[} F^{a}_{4,0}(t) - F^{a}_{6,0}(t)\bigg{]} \cr
&-\frac{t}{15m^{2}}\bigg{[} F^{a}_{4,0}(t) + F^{a}_{4,1}(t) + 5 F^{a}_{5,0}(t)\bigg{]} +\frac{t^{2}}{60m^{4}} F^{a}_{4,1}(t), \cr
\mathcal{J}^{a}_{3}(t) &=-\frac{1}{6} \bigg{[} F^{a}_{4,0}(t) + F^{a}_{4,1}(t)\bigg{]}+\frac{t}{24m^{2}} F^{a}_{4,1}(t), 
\end{align}
\begin{align}
D^{a}_{0}(t) &= F^{a}_{2,0}(t) - \frac{16}{3}F^{a}_{5,0}(t) \cr
&-\frac{t}{6m^{2}}\bigg{[} F^{a}_{2,0}(t) +F^{a}_{2,1}(t) -4F^{a}_{5,0}(t)\bigg{]} +\frac{t^{2}}{24m^{4}}F^{a}_{2,1}(t), \cr
D^{a}_{2}(t) &=\frac{4}{3} F^{a}_{5,0}(t), \cr
D^{a}_{3}(t) &= \frac{1}{6}\bigg{[}-F^{a}_{2,0}(t) -F^{a}_{2,1}(t) + 4 F^{a}_{5,0}(t)\bigg{]}+\frac{t}{24m^{2}}F^{a}_{2,1}(t).
\end{align}
One can refer to Appendix~\ref{Appendix:A} in detail. The sum of the quark and gluon contributions to the GMFFs is also scale-invariant:
\begin{align}
\mathcal{E}_{0,2}(t)=\sum_{a}\mathcal{E}^{a}_{0,2}(t), \ \ \ \mathcal{J}_{1,3}(t)=\sum_{a}\mathcal{J}^{a}_{1,3}(t), \ \ \ D_{0,2,3}(t)=\sum_{a}D^{a}_{0,2,3}(t).
\end{align}
The EMT density $T^{\mu\nu}(\bm{r},\sigma',\sigma)$ is given by the Fourier transform of the matrix element of the EMT current in momentum space:
\begin{align}
T^{\mu\nu}(\bm{r},\sigma',\sigma)=\sum_{a}T_{a}^{\mu\nu}(\bm{r},\sigma',\sigma)= \sum_{a}\int \frac{d^{3} {\Delta}}{2E(2\pi)^{3}}e^{-i\bm{\Delta} \cdot \bm{r}} \langle p', \sigma' | \hat{T}_{a}^{\mu\nu}(0) | p ,\sigma \rangle.
\end{align}

\subsection{Energy density}
The temporal component of the EMT $T^{00}(\bm{r},\sigma',\sigma)$ is related to the energy density. The mutipole expansion of the energy density is defined by
\begin{align}
T^{00}(\bm{r},\sigma',\sigma)& = \sum_{a}\int \frac{d^{3} {\Delta}}{2E(2\pi)^{3}}e^{-i\bm{\Delta} \cdot \bm{r}} \langle p', \sigma' | \hat{T}_{a}^{00}(0) | p ,\sigma \rangle \cr
& = \varepsilon_{0}(r)\delta_{\sigma'\sigma} + \varepsilon_{2}(r) \hat{Q}_{\sigma'\sigma}^{ij} Y_{2}^{ij}(\Omega_{r}),
\label{eq:e_den}
\end{align}
where the monopole and quadrupole densities $\varepsilon_{0,2}(r)$ are respectively given by
\begin{align}
\varepsilon_{0}(r) =  m \tilde{\mathcal{E}}_{0}(r), \ \ \ \varepsilon_{2}(r) = -\frac{1}{m} r\frac{d}{dr} \frac{1}{r}\frac{d}{dr} \tilde{\mathcal{E}}_{2}(r),
\end{align}
with
\begin{align}
 \tilde{\mathcal{E}}_{0,2}(r) = \int \frac{d^{3} {\Delta}}{(2\pi)^{3}}e^{-i\bm{\Delta} \cdot \bm{r}} \mathcal{E}_{0,2}(t).
\end{align}
At the same time, the energy multipole form factors $\mathcal{E}_{0,2}(t)$ can be expressed in terms of the energy densities $\varepsilon_{0,2}(r)$ in coordinate space:
\begin{align}
\mathcal{E}_{0}(t) &= \frac{1}{m}\int d^{3}r \, j_{0}(r\sqrt{-t}) \varepsilon_{0}(r), \cr
\mathcal{E}_{2}(t) &= m\int d^{3}r \, \frac{j_{2}(r\sqrt{-t})}{t} \varepsilon_{2}(r). 
\label{eq:EMTFF_E}
\end{align}
For a particle of arbitrary spin, the general tensor quantity is introduced in Ref.~\cite{Polyakov:2018rew, Cosyn:2019aio} by
\begin{align}
M^{k_{1}...k_{n}}_{n}(\sigma',\sigma) = \sum_{a}\int d^{3}r \, r^n Y^{k_{1}...k_{n}}_{n} T_{a}^{00}(\bm{r},\sigma',\sigma).
\label{eq:multipole}
\end{align}
The monopole moment corresponds to the mass of a baryon, accordingly one arrives at the apparent relation
\begin{align}
M_{0}(\sigma',\sigma)=\sum_{a}\int d^{3} r \ Y_{0}(\Omega_{r}) T_{a}^{00}(\bm{r},\sigma',\sigma) =\int d^{3} r \ \varepsilon_{0}(r)\delta_{\sigma'\sigma} =  m F_{1,0}(0) \delta_{\sigma'\sigma},
\end{align}
which gives the normalization
\begin{align}
 F_{1,0}(0)= \sum_{a} F^{a}_{1,0}(0) = 1.
 \label{eq:energy_constraint}
\end{align}
The constraint $F_{1,0}(0)=1$ for the spin-3/2 baryon coincides with that for the spin-1/2 and spin-1 hadrons. The gravitational quadrupole density of a baryon presents how the energy density is deformed from the spherically symmetric shape. This quantity does not appear in the spherically symmetric hadrons. It can be quantitatively estimated as
\begin{align}
\mathcal{Q}^{ij}_{\sigma'\sigma}&=M_{2}^{ij}(\sigma',\sigma) =\sum_{a} \int d^{3}r~r^{2} Y_{2}^{ij} T_{a}^{00}(\bm{r},\sigma',\sigma) =\frac{2}{15} \hat{Q}^{ij}_{\sigma'\sigma} \int d^{3}r~r^{2} \varepsilon_{2}(r)   \cr
&=-\frac{2}{m} \hat{Q}^{ij}_{\sigma'\sigma} \mathcal{E}_{2}(0) = \frac{1}{3m} \bigg{[} F_{1,0}(0) + F_{1,1}(t)  -4 F_{5,0}(0) \bigg{]} \hat{Q}^{ij}_{\sigma'\sigma}= \frac{1}{3m} \bigg{[} 1 + F_{1,1}(0)  -4 F_{5,0}(0) \bigg{]} \hat{Q}^{ij}_{\sigma'\sigma}.
\label{eq:mass_moment}
\end{align}
Another interesting property is the mass radius of a baryon. It can be derived by the $r^{2}$-weighted energy density in the Breit frame. The expression of the mass radius is found to be
\begin{align}
\langle r^{2}_{E} \rangle &= \frac{\sum_{a}\int d^{3}r~r^{2} T_{a}^{00}(\bm{r})}{\sum_{a}\int d^{3}r  T_{a}^{00}(\bm{r})} = \frac{1}{m} \int d^{3}r~r^{2} \varepsilon_{0}(r) \cr
&= 6 \mathcal{E}'_{0}(0) = 6 F'_{1,0}(0) +  \frac{1}{m^{2}}\bigg{[}-\frac{5}{2}F_{1,0}(t) - F_{1,1}(t) -\frac{3}{2} F_{2,0}(t) +4 F_{5,0}(t) + 3F_{4,0}(t)  \bigg{]}_{t=0}.
\label{eq:mass_radius}
\end{align}

\subsection{Angular momentum density}
The spin  density is given by
\begin{align}
J^{i}(\bm{r},\sigma',\sigma)&=\sum_{a}J^{i}_{a}(\bm{r},\sigma',\sigma)=\epsilon^{ijk}r^{j}\sum_{a}T_{a}^{0k}(\bm{r},\sigma',\sigma) \cr
&= 2\hat{S}^{j}_{\sigma'\sigma}\int \frac{d^{3}\Delta}{(2\pi)^{3}} e^{-i\bm{\Delta}\cdot \bm{r}} \left[ \left(\mathcal{J}_{1}(t)+ \frac{2}{3}t \frac{d\mathcal{J}_{1}(t)}{dt} \right)Y_{0}\delta^{ij} -  t\frac{d \mathcal{J}_{1}(t)}{dt} Y_{2}^{ij} \right]\cr
& +\frac{2}{m^{2}} \hat{O}^{jmn}_{\sigma'\sigma}\int \frac{d^{3}\Delta}{(2\pi)^{3}} e^{-i\bm{\Delta}\cdot \bm{r}} \bigg{[}  t^{2} \frac{d \mathcal{J}_{3}(t)}{dt} Y^{imnj}_{4}  -\left(2t\mathcal{J}_{3}(t) + \frac{4}{7} t^{2} \frac{d \mathcal{J}_{3}(t)}{dt}\right) \delta^{ij}Y^{mn}_{2}  \bigg{]}.
\label{eq:angular}
\end{align}
The angular momentum density is obtained from the $0k$-components of the EMT, which is decomposed into the $0$-, $2$- and $4$-multipole components (see also Ref.~\cite{Lorce:2017wkb, Schweitzer:2019kkd}). 
The sum of the angular momentum contributions from quark and gluons to the spin-3/2 baryon is obtained by
\begin{align}
J^{i}_{\sigma'\sigma}=\sum_{a}\int d^{3} r \, J^{i}_{a}(\bm{r},\sigma',\sigma)=2\mathcal{J}_{1}(0)\hat{S}^{i}_{\sigma'\sigma} =\frac{2}{3} F_{4,0}(0)\hat{S}^{i}_{\sigma' \sigma},
\label{eq:rhoJ}
\end{align}
which yields the spin operator of the baryon with the constraint $F_{4,0}(0)=3/2$.

Since our interest lies in the monopole angular momentum density in this work, we separately define it as
\begin{align}
J^{i}_{{\mathrm{mono}}}(\bm{r},\sigma',\sigma) = 2\hat{S}^{j}_{\sigma'\sigma}\int \frac{d^{3}\Delta}{(2\pi)^{3}} e^{-i\bm{\Delta}\cdot \bm{r}} \left[ \left(\mathcal{J}_{1}(t)+ \frac{2}{3}t \frac{d\mathcal{J}_{1}(t)}{dt} \right)Y_{0}\delta^{ij} \right],
\end{align}
accordingly the averaged angular momentum density is given by
\begin{align}
\frac{1}{\mathrm{Tr}[\hat{\bm{S}}^{2}]}\sum_{\sigma',\sigma,i}\hat{S}^{i}_{\sigma\sigma'} J^{i}_{\mathrm{mono}}(\bm{r},\sigma',\sigma)= \rho_{J}(r)/S, \ \ \ \mathrm{i.e,}  \ \ \ \rho_{J}(r) = -r \frac{d}{dr} \int \frac{d^{3}\Delta}{(2\pi)^{3}} e^{-\bm{\Delta \cdot \bm{r}}}\mathcal{J}_{1}(t),
\label{eq:a_den}
\end{align}
with spin $S=3/2$.
The angular momentum form factor can be expressed in terms of the averaged angular momentum density as
\begin{align}
\mathcal{J}_{1}(t) &= \int d^{3}r \, \frac{j_{1}(r\sqrt{-t})}{r\sqrt{-t}} \rho_{J}(r).
\label{eq:EMTFF_J}
\end{align}

\subsection{Pressure and shear force densities}
The pressure and shear force densities are related to the $ij$-components of the static EMT. These densities are firstly defined in Ref.~\cite{Polyakov:2019lbq, Polyakov:2018rew} and newly parametrized in Ref.~\cite{Panteleeva:2020ejw, Sun:2020wfo} to conveniently express the strong forces in a hadron acting on the radial area element. Following Ref.~\cite{Panteleeva:2020ejw, Sun:2020wfo} we can express the stress tensor in terms of the pressure and shear forces densities by
\begin{align}
T^{ij}(\bm{r},\sigma',\sigma) &= \sum_{a}\int \frac{d^{3} \Delta}{2E(2\pi)^{3}}  e^{-i \bm{\Delta} \cdot \bm{r}} \langle p', \sigma' | \hat{T}_{a}^{ij}(0) |  p,\sigma \rangle \cr
&={p}_{0}(r) \delta^{ij} \delta_{\sigma'\sigma} + {s}_{0}(r) Y^{ij}_{2}\delta_{\sigma'\sigma} + \left({p}_{2}(r) + \frac{1}{3}p_{3}(r) - \frac{1}{9}s_{3}(r) \right)\hat{Q}^{ij}_{\sigma'\sigma} \cr 
&+ \left({s}_{2}(r)-\frac{1}{2}p_{3}(r)+\frac{1}{6}s_{3}(r)\right)2\left[ \hat{Q}^{ip}_{\sigma'\sigma}Y_{2}^{pj} + \hat{Q}^{jp}_{\sigma'\sigma}Y_{2}^{pi} -\delta^{ij}\hat{Q}^{pq}_{\sigma'\sigma}Y_{2}^{pq}\right] \cr
&+ \hat{Q}^{pq}_{\sigma'\sigma}Y^{pq}_{2}\left[ \left( \frac23 p_{3}(r) + \frac19 s_{3}(r)\right) \delta^{ij} + \left( \frac12 p_{3} (r) + \frac56 s_{3}(r) \right) Y^{ij}_{2}\right].
\label{eq:m_den}
\end{align}
From the EMT conservation $\partial_{i} T^{ij}(\bm{r},\sigma',\sigma) = 0$, the following equilibrium relations between the pressure and shear force densities are derived:
\begin{align}
\frac{2}{3}\frac{d s_{n}(r)}{dr} + 2 \frac{s_{n}(r)}{r} +\frac{d p_{n}(r)}{dr} = 0, \ \ \mathrm{ with }  \ \ n=0,2,3.
\label{eq:EMT_differential}
\end{align}
This differential equation guarantees the stability condition. The functions $p_{0}(r)$ and $s_{0}(r)$ correspond to the pressure and shear force densities appearing in the spherically symmetric hadrons. The functions $p_{2}(r)$ and $p_{3}(r)$ are named the quadrupole pressure densities, and the $s_{2}(r)$ and $s_{3}(r)$ are called the quadrupole shear force densities according to Ref.~\cite{Panteleeva:2020ejw}.
These densities $p_{n}(r)$ and $s_{n}(r)$ are respectively written as
\begin{align}
&p_{n}(r)= \frac{1}{6m} \partial^{2} \tilde{\mathcal{D}}_{n}(r) = \frac{1}{6m} \frac{1}{r^{2}}\frac{d}{dr} r^{2} \frac{d}{dr}\tilde{\mathcal{D}}_{n}(r), \cr
&s_{n}(r)= -\frac{1}{4m} r \frac{d}{dr} \frac{1}{r} \frac{d}{dr} \tilde{\mathcal{D}}_{n}(r),
\label{eq:ps_FT}
\end{align}
with\footnote{a typo in Ref.~\cite{Sun:2020wfo} is corrected in Eq.~\eqref{eq:FF_dis}.}
\begin{align}
&\tilde{\mathcal{D}}_{0}(r) = \int \frac{d^{3}\Delta}{(2\pi)^{3}} e^{-i\bm{\Delta}\cdot \bm{r}} D_{0}(t),  \cr
&\tilde{\mathcal{D}}_{2}(r) = \int \frac{d^{3}\Delta}{(2\pi)^{3}} e^{-i\bm{\Delta}\cdot \bm{r}} D_{2}(t) + \frac{1}{m^{2}} \left(\frac{d}{dr}\frac{d}{dr} - \frac{2}{r} \frac{d}{dr}\right) \int \frac{d^{3}\Delta}{(2\pi)^{3}} e^{-i\bm{\Delta}\cdot \bm{r}} D_{3}(t), \cr
&\tilde{\mathcal{D}}_{3}(r) = -\frac{2}{m^{2}} \left(\frac{d}{dr}\frac{d}{dr} - \frac{3}{r} \frac{d}{dr}\right) \int \frac{d^{3}\Delta}{(2\pi)^{3}} e^{-i\bm{\Delta}\cdot \bm{r}} D_{3}(t).
\label{eq:FF_dis}
\end{align}
Similarly, the form factors $D_{0,2,3,}(t)$ can be expressed in terms of the pressure and shear force  densities in coordinate space:
\begin{align}
D_{0}(t) &= 6m\int d^{3}r \, \frac{j_{0}(r\sqrt{-t})}{t} p_{0}(r), \cr
D_{2}(t) &= 2m\int d^{3}r \, \frac{j_{2}(r\sqrt{-t})}{t} \left(2s_{2}(r)-\frac{1}{2}p_{3}(r)+\frac{2}{3}s_{3}(r)\right), \cr
D_{3}(t) &= 4m^{3}\int d^{3}r \, \frac{j_{4}(r\sqrt{-t})}{t^{2}} \left(\frac{1}{2}p_{3}(r)+\frac{5}{6}s_{3}(r)\right).
\label{eq:EMTFF_D}
\end{align}
The pressure densities $p_{n}(r)$ satisfying the relation given in Eq.~\eqref{eq:ps_FT} comply with the von Laue condition
\begin{align}
 \int d^{3}r \, p_{n}(r) = \frac{1}{6m} \int d^{3}r \, \partial^{2} \tilde{\mathcal{D}}_{n}(r)=0. \ \ \   \mathrm{with}\ \ n=0,2,3.
\label{eq:von_Laue}
\end{align}
Note that the dimensionless constants (generalized $D$-terms) are defined by~\cite{Panteleeva:2020ejw}
\begin{align}
\mathcal{D}_{n} \equiv  \int d^{3}r \, \tilde{\mathcal{D}}_{n}(r)= m \int d^{3}r \, r^{2} p_{n}(r)=-\frac{4}{15} m \int d^{3}r \, r^{2} s_{n}(r), \ \ \   \mathrm{with}\ \ n=0,2,3.
\label{eq:D_term}
\end{align}
The generalized $D$-terms $\mathcal{D}_{0,2,3}$ introduced in Ref.~\cite{Panteleeva:2020ejw} are related to the form factors $D_{0,2,3}(t)$ as follows:
\begin{align}
\mathcal{D}_{0}=D_{0}(0), \ \ \ \mathcal{D}_{2}=D_{2}(0)+\frac{2}{m^{2}}\int^{0}_{-\infty} dt \, D_{3}(t), \ \ \ \mathcal{D}_{3}=-\frac{5}{m^{2}}\int^{0}_{-\infty} dt \, D_{3}(t).
\label{eq:D_rel}
\end{align}
Interestingly, the strong forces carried by constituents can be interpreted as a certain combination of pressure and shear force densities~\cite{Polyakov:2002yz}. The spherical components of the strong forces ($dF_{r},dF_{\theta}$ and $dF_{\phi}$)  acting on the radial area element ($d\bm{S}=dS_{r}\hat{\bm{e}}_{r}+dS_{\theta}\hat{\bm{e}}_{\theta}+dS_{\phi}\hat{\bm{e}}_{\phi}$) are expressed as follows~\cite{Panteleeva:2020ejw}:
\begin{align}
\frac{dF_{r}}{dS_{r}}&=\delta_{\sigma'\sigma}\left(p_{0}(r)+\frac{2}{3}s_{0}(r)\right) + \hat{Q}_{\sigma'\sigma}^{rr}\left(p_{2}(r)+\frac{2}{3}s_{2}(r)+p_{3}(r)+\frac{2}{3}s_{3}(r)\right), \cr
\frac{dF_{\theta}}{dS_{r}}&=\hat{Q}^{\theta r}_{\sigma'\sigma}\left(p_{2}(r)+\frac{2}{3}s_{2}(r)\right), \ \ \ \ \ \frac{dF_{\phi}}{dS_{r}}=\hat{Q}_{\sigma'\sigma}^{\phi r}\left(p_{2}(r)+\frac{2}{3}s_{2}(r)\right).
\label{eq:force}
\end{align}
Here, as defined in Ref.~\cite{Polyakov:2018zvc}, the mechanical radius can be given by
\begin{align}
\langle r_{n}^{2} \rangle_{\mathrm{mech}} = \frac{\int d^{3}r \, r^{2} \left[p_{n}(r) + \frac{2}{3}s_{n}(r)\right]}{\int d^{3}r \, \left[p_{n}(r) + \frac{2}{3}s_{n}(r)\right]}.
\label{eq:mech}
\end{align}
As for the unpolarized spin-3/2 hadron, since the normal force acting on the radial area element (${dF_{r}}/{dS_{r}}$) is solely due to $p_{0}(r)+\frac{2}{3}s_{0}(r)$, it should comply with the local stability criterion given in Ref.~\cite{Polyakov:2018zvc} as
\begin{align}
p_{0}(r) + \frac{2}{3} s_{0}(r) > 0.
\label{eq:local}
\end{align}
\section{Gravitational form factors of the $\Delta$ in the Skyrme model \label{sec:3}}
The Skyrme Lagrangian density is given by
\begin{align}
\mathcal{L}= \frac{F^{2}_{\pi}}{16}\mathrm{tr}_{F}\left[\partial_{\mu} U\partial^{\mu}U^{\dagger} \right] + \frac{1}{32e^{2}}\mathrm{tr}_{F}\left[(\partial_{\mu} U)U^{\dagger},(\partial_{\nu} U)U^{\dagger} \right]^{2} + \frac{m^{2}_{\pi}F^{2}_{\pi}}{8}\mathrm{tr}_{F}\left[U-1\right],
\end{align}
where $U$ is the SU(2) chiral field, and $e$ stands for a dimensionless parameter and $\mathrm{tr}_{F}$ is trace over flavors. The $F_{\pi}$ and the $m_{\pi}$ are the pion decay constant and the pion mass, respectively. 

In the large-$N_{c}$ limit, we have the parameter scales $F_{\pi} = \mathcal{O}(N^{1/2}_{c}), \, e = \mathcal{O}(N^{-1/2}_{c}), \, m_{\pi} = \mathcal{O}(N^{0}_{c}),$ so as to have $\mathcal{L}=\mathcal{O}(N_{c})$.
In this limit, the chiral field is assumed to be a static one. Here, the static chiral field is written as $U(r)=\mathrm{exp}[\hat{r}^{i}\tau^{i} P(r)]$, with $\hat{r}^{i}={r^{i}}/{|\bm{r}|}$ and the isospin Pauli matrices $\tau^{i}$, by adopting the hedgehog ansatz where $P(r)$ indicates a profile function with the boundary conditions $P(0)\to \pi$ and $P(\infty) \to 0$.

The classical soliton mass is defined by $M_{\mathrm{sol}}=-\int d^{3}r \, \mathcal{L}$, which is expressed with respect to the profile function and the model parameters by
\begin{align}
M_{\mathrm{sol}}[P]=4\pi \int^{\infty}_{0}dr \,r^{2} \left[\frac{F^{2}_{\pi}}{8}\left(\frac{2\mathrm{sin}^{2}P(r)}{r^{2}} + P'(r)^{2}\right) + \frac{\mathrm{sin}^{2}P(r)}{2e^{2}r^{2}}\left(\frac{\mathrm{sin}^{2}P(r)}{r^{2}} + 2P'(r)^{2}\right) + \frac{m^{2}_{\pi}F^{2}_{\pi}}{4}\left(1-\mathrm{cos}P(r)\right)\right].
\end{align}
The profile function is obtained by minimizing the classical soliton mass and has an asymptotic behavior in the limiting case $r\to \infty$,
\begin{align}
P(r) = \frac{2R^{2}_{0}}{r^{2}}(1+m_{\pi}r)e^{-m_{\pi}r}  \, \, \, (\mathrm{large} \ r),
\label{eq:profile}
\end{align}
where the constant $R_{0}$ is determined by the profile function derived by minimizing the classical soliton mass and is given in terms of the axial coupling constant in the chiral limit~\cite{Cebulla:2007ei, Perevalova:2016dln}.

In order to assign the quantum number to the soliton, the time-dependent chiral field should be considered as $U(r)\to A(t)U(r)A^{\dagger}(t)$, with $A(t)=a_{0}+\bm{a}\cdot\tau$. Here, we define the angular velocity $\Omega(t)= \dot{A}^{\dagger}A$ ($\dot{A}\equiv dA/dt$) with 
\begin{align}
\Omega^{i}=-\frac{i}{2}\mathrm{tr}_{F}[\dot{A}^{\dagger}A \tau^{i}].
\end{align}
 Having quantized the collective coordinates ($\Omega^{i} \to \hat{J}^{i}/2I$), one obtains the collective Hamiltonian~\cite{Adkins:1983ya} 
\begin{align}
H= M_{\mathrm{sol}} + \frac{\hat{\bm{J}}^{2}}{2I},
\end{align}
with the moment of inertia 
\begin{align}
I= \frac{2\pi}{3} \int dr \, r^{2} \mathrm{sin}P(r) \left[ F_{\pi}^{2} + \frac{4P'(r)^{2}}{e^{2}} + \frac{4\mathrm{sin}^{2}P(r)}{e^{2}r^{2}} \right].
\end{align}
The collective Hamiltonian acts on the collective baryon wave functions given in Ref.~\cite{Adkins:1983ya}. In principle $F_{\pi}$ and $e$ are model parameters. However, the parameters are fixed to repoduce the following observables~\cite{Cebulla:2007ei}
\begin{align}
M_{N}+M_{\Delta}\equiv2M_{\mathrm{sol}}=2171 \,\mathrm{MeV}, \ \ \ \ \ \ M_{\Delta}-M_{N}\equiv\frac{3}{2I}= 293 \,\mathrm{MeV},
\end{align}
and the model parameters are consequently found to be
\begin{align}
m_{\pi}=138 \, \mathrm{MeV}, \ \ \ F_{\pi}=131.3 \, \mathrm{MeV}, \ \ \ e=4.628.
\end{align}
From the above parameters, the classical soliton mass and the moment of inertia are determined by
\begin{align}
M_{\mathrm{sol}} = 1085 \, \mathrm{MeV}, \ \ \ I = 1.01 \, \mathrm{fm}.
\end{align}
In this work, we strictly follow the set of parameters used in Ref.~\cite{Cebulla:2007ei} to keep consistency.

The canonical EMT in the Skyrme model can be derived by
\begin{align}
T^{\mu \nu} = \frac{\partial \mathcal{L}}{\partial (\partial_{\mu} \phi_{a})} \partial^{\nu} \phi_{a} - g^{\mu \nu} \mathcal{L},
\end{align}
where $\phi_{a}$ is a time-dependent mesonic field with $U(t,r) = \phi_{0}+ i \bm{\tau}\cdot\bm{\phi}$. The degree of freedom of the mesonic field is reduced to three $(a=1,2,3)$ by the constraint $\phi^{2}_{0} + \bm{\phi}^{2}=1$. With the constraint, the canonical EMT in the Skyrme model is found to be symmetric. The respective components of the EMT densities is expressed as 
\begin{align}
T^{00} (\bm{r},\sigma',\sigma) &= \delta_{\sigma'\sigma}\left[\frac{F^{2}_{\pi}}{8}\left(\frac{2\mathrm{sin}^{2}P(r)}{r^{2}} + P'(r)^{2}\right) + \frac{\mathrm{sin}^{2}P(r)}{2e^{2}r^{2}}\left(\frac{\mathrm{sin}^{2}P(r)}{r^{2}} + 2P'(r)^{2}\right) + \frac{m^{2}_{\pi}F^{2}_{\pi}}{4}\left(1-\mathrm{cos}P(r)\right)\right],\cr
T^{ij} (\bm{r},\sigma',\sigma) &= \hat{r}^{i}\hat{r}^{j}\delta_{\sigma'\sigma}\left[\frac{F^{2}_{\pi}}{8}\left( 2 P'(r)^{2} - \frac{2\mathrm{sin}^{2}P(r)}{r^{2}} \right) + \frac{\mathrm{sin}^{2}P(r)}{e^{2}r^{2}}\left(P'(r)^{2}-\frac{\mathrm{sin}^{2}P(r)}{r^{2}} \right)\right] \cr
& \hspace{0.3cm}+ \delta^{ij}\delta_{\sigma'\sigma}\left[-\frac{F_{\pi}^{2}}{8}P'(r)^{2} + \frac{\mathrm{sin}^{4}P(r)}{2e^{2}r^{4}} - \frac{m^{2}_{\pi}F^{2}_{\pi}}{4}\left(1-\mathrm{cos}P(r)\right)\right],\cr
T^{0k} (\bm{r},\sigma',\sigma) &= (\hat{\bm{J}}\times\hat{\bm{r}})^{k}_{\sigma'\sigma} \frac{\mathrm{sin}^{2}P(r)}{4Ir} \left[F_{\pi}^{2} +\frac{4\mathrm{sin}^{2}P(r)}{e^{2}r^{2}} + \frac{4P'(r)^{2}}{e^{2}}\right].
\label{eq:EMT_leading}
\end{align}
The rotational corrections to the EMT densities are obtained by
\begin{align}
\delta_{\mathrm{rot}} T^{00} (\bm{r},\sigma',\sigma) &=\left[\hat{\bm{J}}^{2}-(\hat{\bm{J}}\cdot\hat{\bm{r}})^{2}\right]_{\sigma' \sigma}\frac{\mathrm{sin}^{2}P(r)}{8I^{2}}  \left[F^{2}_{\pi} + \frac{4P'(r)^{2}}{e^{2}} + \frac{4\mathrm{sin}^{2}P(r)}{e^{2}r^{2}} \right], \cr
\delta_{\mathrm{rot}} T^{0k} (\bm{r},\sigma',\sigma) &= \mathcal{O}(1/N^{2}_{c}), \cr
\delta_{\mathrm{rot}} T^{ij} (\bm{r},\sigma',\sigma) &=\hat{r}^{i}\hat{r}^{j}\left[\hat{\bm{J}}^{2}-(\hat{\bm{J}}\cdot\hat{\bm{r}})^{2}\right]_{\sigma' \sigma}\frac{\mathrm{sin}^{2}P(r)}{8I^{2}}  \left[ -\frac{8P'(r)^{2}}{e^{2}} + \frac{8\mathrm{sin}^{2}P(r)}{e^{2}r^{2}} \right] \cr
&\hspace{0.2cm} + \delta^{ij} \left[\hat{\bm{J}}^{2}-(\hat{\bm{J}}\cdot\hat{\bm{r}})^{2}
\right]_{\sigma' \sigma}\frac{\mathrm{sin}^{2}P(r)}{8I^{2}}  \left[F^{2}_{\pi} + \frac{4P'(r)^{2}}{e^{2}} + \frac{4\mathrm{sin}^{2}P(r)}{e^{2}r^{2}} \right] \cr
&\hspace{0.2cm} -\left[2\hat{\bm{J}}^{2}\hat{r}^{i}\hat{r}^{j}-\hat{r}^{i}\hat{r}^{l}\{\hat{J}^{l},\hat{J}^{j} \}-\hat{r}^{j}\hat{r}^{l}\{\hat{J}^{l},\hat{J}^{i}\} + \{\hat{J}^{i},\hat{J}^{j}\} \right]_{\sigma' \sigma} \frac{\mathrm{sin}^{4}P(r)}{2I^{2}e^{2}r^{2}}.
\label{eq:EMT_rot}
\end{align}
The rotational corrections to the ${0k}$-components of the EMT densities $\delta_{\mathrm{rot}}T^{0k}(\bm{r},\sigma',\sigma)$ are found to be null in the current Skyrme Lagrangian. The corrections appear in the higher-order derivative terms that generate $\Omega^{3}\sim\mathcal{O}(1/N_{c}^{3})$. However, the corrections are strongly suppressed in the large-$N_{c}$ expansion. Thus, we can safely neglect the rotational corrections to the $0k$-component of the EMT densities.

As given in Eqs.~\eqref{eq:e_den},~\eqref{eq:a_den} and~\eqref{eq:m_den}, the multipole densities can be extracted from the EMT densities in Eq.~\eqref{eq:EMT_leading}:
\begin{align}
\varepsilon_{0}(r) &= \left[\frac{F^{2}_{\pi}}{8}\left(\frac{2\mathrm{sin}^{2}P(r)}{r^{2}} + P'(r)^{2}\right) + \frac{\mathrm{sin}^{2}P(r)}{2e^{2}r^{2}}\left(\frac{\mathrm{sin}^{2}P(r)}{r^{2}} + 2P'(r)^{2}\right) + \frac{m^{2}_{\pi}F^{2}_{\pi}}{4}\left(1-\mathrm{cos}P(r)\right)\right], \cr
\rho_{J}(r)&= \frac{\mathrm{sin}^{2}P(r)}{4I} \left[F_{\pi}^{2} +\frac{4\mathrm{sin}^{2}P(r)}{e^{2}r^{2}} + \frac{4P'(r)^{2}}{e^{2}}\right], \cr
p_{0}(r) &= -\frac{F^{2}_{\pi}}{24}\left(\frac{2\mathrm{sin}^{2}P(r)}{r^{2}} + P'(r)^{2}\right) + \frac{\mathrm{sin}^{2}P(r)}{6e^{2}r^{2}}\left(\frac{\mathrm{sin}^{2}P(r)}{r^{2}} + 2P'(r)^{2}\right) - \frac{m^{2}_{\pi}F^{2}_{\pi}}{4}\left(1-\mathrm{cos}P(r)\right), \cr
s_{0}(r) &= \left(\frac{F^{2}_{\pi}}{4}+\frac{\mathrm{sin}^{2}P(r)}{e^{2}r^{2}}\right)\left(P'(r)^{2}-\frac{\mathrm{sin}^{2}P(r)}{r^{2}} \right).
\end{align}
In the same manner, the rotational corrections are derived from Eq.~\eqref{eq:EMT_rot}: 
\begin{align}
\delta_{\mathrm{rot}}\varepsilon^{(J)}_{0}(r) &=  J(J+1) \frac{\mathrm{sin}^{2}P(r)}{12I^{2}}  \left[F^{2}_{\pi} + \frac{4P'(r)^{2}}{e^{2}} + \frac{4\mathrm{sin}^{2}P(r)}{e^{2}r^{2}} \right], \cr
\delta_{\mathrm{rot}}\rho^{(J)}_{J}(r) &=\mathcal{O}(1/N^{2}_{c}), \cr
\delta_{\mathrm{rot}}p^{(J)}_{0}(r) &=  J(J+1) \frac{\mathrm{sin}^{2}P(r)}{12I^{2}}  \left[F^{2}_{\pi} + \frac{4P'(r)^{2}}{3e^{2}} + \frac{4\mathrm{sin}^{2}P(r)}{3e^{2}r^{2}} \right], \cr
\delta_{\mathrm{rot}}s^{(J)}_{0}(r) &=  J(J+1) \frac{\mathrm{sin}^{2}P(r)}{12I^{2}}  \left[ - \frac{8P'(r)^{2}}{e^{2}} + \frac{4\mathrm{sin}^{2}P(r)}{e^{2}r^{2}} \right].
\label{eq:rot_density}
\end{align}
Interestingly, in the chiral soliton picture the rotational corrections to the EMT densities in Eq.~\eqref{eq:rot_density} are related to the quadrupole densities, which was firstly found in Ref.~\cite{Panteleeva:2020ejw}. The quadrupole energy density $\varepsilon_{2}(r)$ is found to be
\begin{align}
\delta_{\mathrm{rot}}\varepsilon^{(J)}_{0}(r)=-\frac{2}{3}J(J+1)\varepsilon_{2}(r).
\label{eq:rot_energy} 
\end{align}
In other words, the quadrupole energy density $\varepsilon_{2}(r)$ is related to the energy densities of the nucleon $\varepsilon^{N}_{0}(r)$ and the $\Delta$ $\varepsilon^{\Delta}_{0}(r)$
\begin{align}
\varepsilon_{2}(r)=-\frac{1}{2}\left[\varepsilon^{\Delta}_{0}(r) -\varepsilon^{N}_{0}(r)\right],
\label{eq:NdeltaE2}
\end{align}
with
\begin{align}
\varepsilon^{N,\Delta}_{0}(r)\equiv\left[\varepsilon_{0}(r) + \delta_{\mathrm{rot}}\varepsilon^{(\frac12,\frac32)}_{0}(r)\right].
\end{align}
Also, the usual sizes of the quadrupole density $\varepsilon_{2}(r)$ is estimated by
\begin{align}
\int d^{3} r \, [\varepsilon^{\Delta}_{0}(r) - \varepsilon^{N}_{0}(r)] =\int d^{3} r \, [\delta_{\mathrm{rot}}\varepsilon^{{(\frac32)}}_{0}(r) - \delta_{\mathrm{rot}}\varepsilon^{(\frac12)}_{0}(r)]=-2\int d^{3} r \, \varepsilon_{2}(r) = \frac{3}{2I} \sim \mathcal{O}(1/N_{c}).
\label{eq:energy_Int}
\end{align}

As for the quadrupole pressure and shear force  densities, the remarkable general relation in a chiral soliton picture is derived in Ref.~\cite{Panteleeva:2020ejw} 
\begin{align}
p_{2}(r)+ \frac{2}{3}s_{2}(r) =0.
\label{eq:null}
\end{align}
By comparing Eq.~\eqref{eq:m_den} with Eq.~\eqref{eq:EMT_rot} we reproduce this relation in the Skyrme model. The above relation~\eqref{eq:null}, together with the EMT conservation~\eqref{eq:EMT_differential}, implies the null results of the quadrupole densities $s_{2}(r)$ and $p_{2}(r)$, and the similar relations to Eq.~\eqref{eq:rot_energy} for the pressure and shear force densities are obtained~\cite{Panteleeva:2020ejw}:
\begin{align}
p_{2}(r)=s_{2}(r)=0, \ \ \  \delta_{\mathrm{rot}}p^{(J)}_{0}(r) = -\frac{2}{3} J(J+1) p_{3}(r), \ \ \ \delta_{\mathrm{rot}}s^{(J)}_{0}(r) = -\frac{2}{3} J(J+1) s_{3}(r).
\label{eq:quad_density}
\end{align}
Therefore, we arrive at the similar expressions as Eq.~\eqref{eq:NdeltaE2}
\begin{align}
p_{3}(r)=-\frac{1}{2}\left[p^{\Delta}_{0}(r) -p^{N}_{0}(r)\right], \ \ \ \ s_{3}(r)=-\frac{1}{2}\left[s^{\Delta}_{0}(r) -s^{N}_{0}(r)\right],
\label{eq:Ndeltap3}
\end{align}
with
\begin{align}
p^{N,\Delta}_{0}(r)\equiv\left[p_{0}(r) + \delta_{\mathrm{rot}}p^{(\frac12,\frac32)}_{0}(r)\right], \ \ \ \ \ s^{N,\Delta}_{0}(r)\equiv\left[s_{0}(r) + \delta_{\mathrm{rot}}s^{(\frac12,\frac32)}_{0}(r)\right].
\end{align}
Here, one has to bear in mind that due to the EMT densities with the included rotational corrections the pressure $p_{n}(r)$ and the shear force $s_{n}(r)$ densities do not uniquely exist and should be reasonably determined. Let us first recall that the leading-order (LO) result of $p_{0}(r)$ satisfies the von Laue condition that is equivalent to the equation of motion~\cite{Cebulla:2007ei}. However, once one considers the next-to-leading-order (NLO) result of $p_{0}(r)$, it breaks the von Laue condition. Thus, one might introduce the ``variation after quantization'' method—minimizing the baryon mass after quantizing the soliton—as a prescription. However, this method also has a drawback: the chiral symmetry in the large-$r$ region is not satisfied~\cite{Perevalova:2016dln}. To preserve the chiral symmetry and satisfy the von Laue condition, we first adopt the ``quantization after variation" method—quantizing the soliton after minimizing the soliton mass—and treat the rotational corrections to the EMT densities as a small perturbation. Of course, the von Laue condition is broken. Thus, instead of directly using the model result of $p_{0}(r)$, we solve the equilibrium equation  given in Eq.~\eqref{eq:EMT_differential} with the approximated shear force density $s_{0}(r)$ and reconstruct the pressure $p_{0}(r)|_{\mathrm{reconst}}$, which then automatically complies with the stability condition~\cite{Perevalova:2016dln}. We also derive the reconstructed quadrupole pressure $p_{3}(r)|_{\mathrm{reconst}}$ in the same manner.

Before discussing the GFFs, it is important to discuss the large distance properties of the EMT densities. The large distance behaviors of the EMT densities for the spherically symmetric baryon were investigated and presented within the Skyrme model in Ref.~\cite{Cebulla:2007ei, Perevalova:2016dln}. For completeness, we present the large distance properties of the quadrupole densities in the chiral limit as
\begin{align}
\varepsilon_{2}(r) = -\frac{F^{2}_{\pi}}{2I^{2}}\frac{R^{4}_{0}}{r^{4}} \cdots, \ p_{3}(r) = -\frac{F^{2}_{\pi}}{2I^{2}}\frac{R^{4}_{0}}{r^{4}} \cdots, \ s_{3}(r) = \frac{56}{I^{2}e^{2}}\frac{R^{8}_{0}}{r^{10}} \cdots, \ p_{3}(r)\bigg{|}_{\mathrm{reconst}} = -\frac{392}{15}\frac{1}{I^{2}e^{2}}\frac{R^{8}_{0}}{r^{10}}\cdots.
\label{eq:chiral_prop}
\end{align}
The $\cdots$ indicates the contributions strongly suppressed in the large-$r$ region. Interestingly, the quadrupole densities $\varepsilon_{2}(r)$ and $p_{3}(r)$ are weakly suppressed in the large distance, which have the analogous behavior with the angular momentum density $\rho_{J}(r)\propto \frac{1}{r^{4}}$. Keep in mind that in order to respect the chiral physics and the stability condition, we discard the result of $p_{3}(r)$ and adopt the newly reconstructed $p_{3}(r)|_{\mathrm{reconst}}$ by solving the differential equation~\eqref{eq:EMT_differential}. As a result, the discrepancy between $p_{3}(r)|_{\mathrm{reconst}}$ and $p_{3}(r)$ arises from the fulfillment of both the chiral physics and the stability condition and is inevitable. For finite pion mass, the densities are exponentially suppressed.

Since this model is based on the large-$N_{c}$ expansion, it is necessary to clarify the large-$N_{c}$ behaviors of the GMFFs of the $\Delta$. The large-$N_c$ expansion is valid in the region $|t|\ll M^{2}_{\Delta}$. In the limit of $N_{c} \to \infty$, we have the following large-$N_{c}$ behaviors
\begin{align}
M_{N,\Delta} \sim \mathcal{O}(N_{c}),  \ \ \ I \sim \mathcal{O}(1/N_{c}),  \ \ \ t \sim \mathcal{O}(N^{0}_{c}),
\end{align}
and the scales of the GMFFs are found to be
\begin{align}
&\mathcal{E}_{0}(t) \sim \mathcal{O}(N^{0}_{c}), \ \ \ \ \ \mathcal{E}_{2}(t) \sim \mathcal{O}(N^{0}_{c}),
\ \ \ \ \ \mathcal{J}_{0}(t) \sim \mathcal{O}(N^{0}_{c}),
\ \ \ \ \ \mathcal{J}_{3}(t) \sim \mathcal{O}(N^{0}_{c}), \cr
&D_{0}(t) \sim \mathcal{O}(N^{2}_{c}), \ \ \ \ D_{2}(t) \sim \mathcal{O}(N^{0}_{c}),
\ \ \ \ D_{3}(t) \sim \mathcal{O}(N^{2}_{c}),
\end{align}
or according to Ref.~\cite{Panteleeva:2020ejw} the generalized $D$-terms have the scales as
\begin{align}
&{\mathcal{D}}_{0} \sim \mathcal{O}(N^{2}_{c}), \ \ \ \ {\mathcal{D}}_{2} \sim \mathcal{O}(N^{0}_{c}),
\ \ \ \ {\mathcal{D}}_{3} \sim \mathcal{O}(N^{0}_{c}).
\end{align}
It is found that the GMFFs have the orders of $\sim N^{0}_{c}$ except for the form factors $D_{0}(t)$ and $D_{3}(t)$, which have the orders of $\sim N^{2}_{c}$, whereas the generalized $D$-terms ${\mathcal{D}}_{0}, {\mathcal{D}}_{2}$ and ${\mathcal{D}_{3}}$ have the orders of $\sim N^{2}_{c}$, $\sim N^{0}_{c}$ and $\sim N^{0}_{c}$, respectively.

\section{numerical results and discussion\label{sec:4}}
In this Section, we present the numerical results and discuss them.
We first examine the monopole and quadrupole energy densities arising from the
temporal component of the EMT. Note that the LO monopole energy densities $\varepsilon_{0}(r)$ of
the nucleon and the $\Delta$ are degenerate. To lift the degeneracy, we need to take into account
the rotational corrections $\Omega^{2}\sim\mathcal{O}(1/N_{c}^{2})$.
Thus, the integrations of the NLO monopole energy densities $\varepsilon^{N,\Delta}_{0}(r)$ over space yield
the masses of the nucleon ($M_{N}=1159 \, \mathrm{MeV}$) and
$\Delta$ ($M_{\Delta}=1452 \, \mathrm{MeV}$). While the Skyrme model produces rather
large values of the baryon masses, they satisfy the constraint given in
Eq.~\eqref{eq:energy_constraint},
\begin{align}
  \frac{1}{M_{\mathrm{sol}}}\int d^{3}r \, \varepsilon_{0}(r) \bigg{|}_{LO}
  = \frac{1}{M_{N,\Delta}}\int d^{3}r \, \varepsilon^{N,\Delta}_{0}(r) \bigg{|}_{NLO,S=\frac12,\frac32}
  = F_{1,0}(0)= 1.
\end{align}

\begin{figure}[htp]
\centering
\includegraphics[scale=0.63]{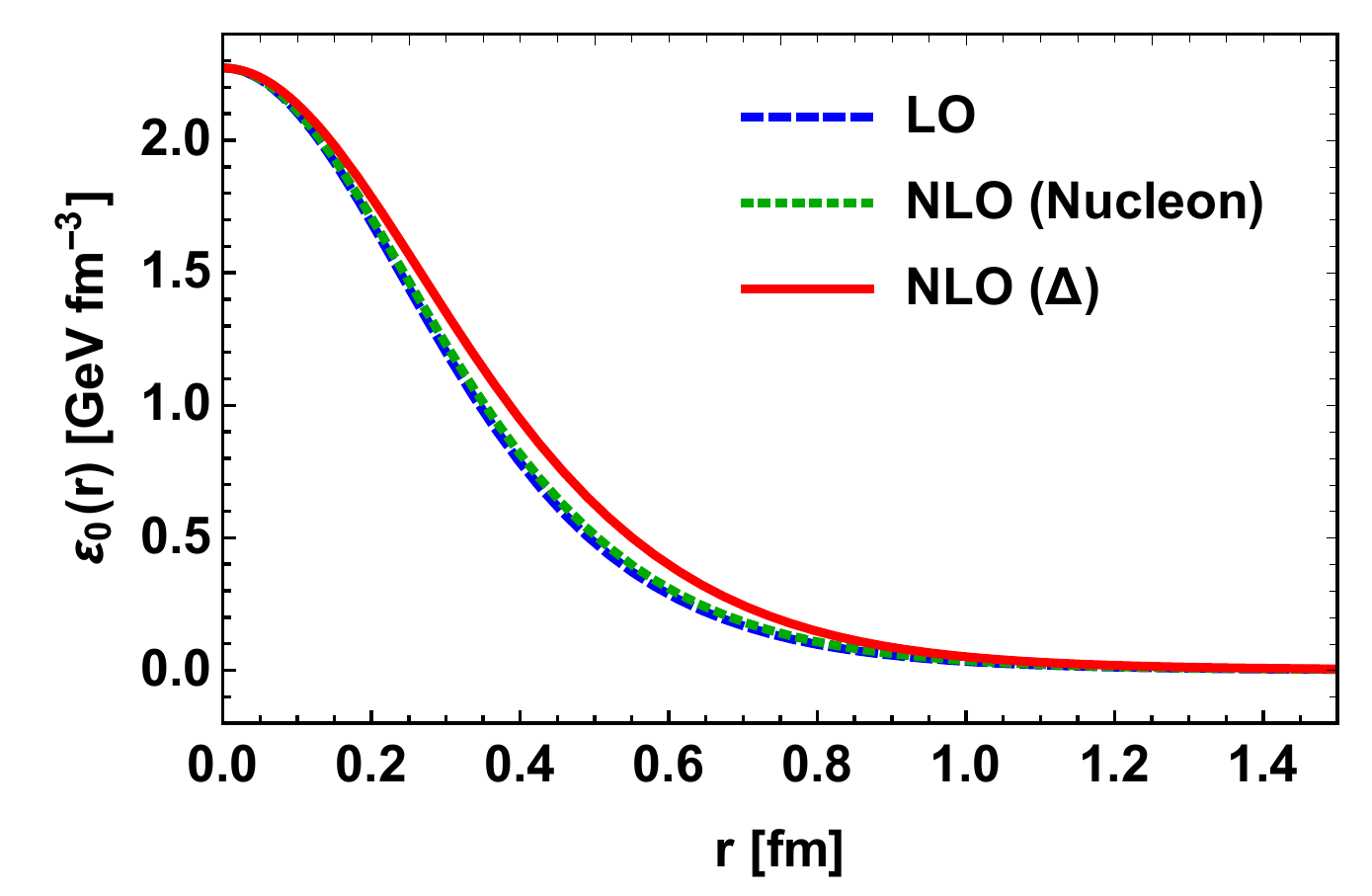}
\includegraphics[scale=0.63]{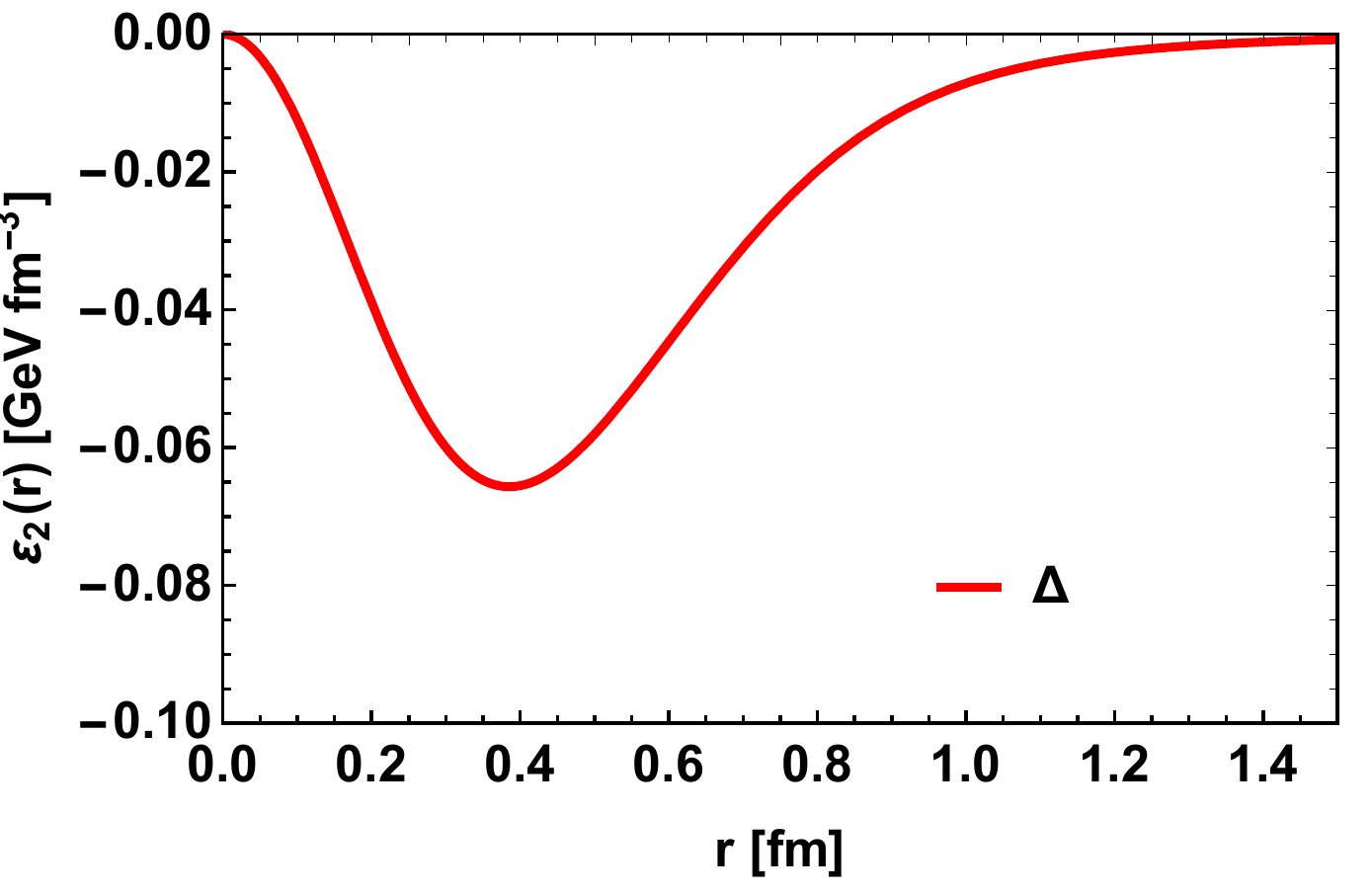}
\caption{The left panel presents the results for the monopole energy densities of the classical soliton (LO), the nucleon (NLO) and the $\Delta$ (NLO) as functions of radius $r$.
  The right panel depicts the quadrupole energy density of the $\Delta$
  as a function of radius $r$. The dashed, dotted and solid curves draw the classical soliton,
  the nucleon and the $\Delta$, respectively. }
\label{fig:1}
\end{figure}
In the left panel of Fig.~\ref{fig:1}, we draw the numerical results on the monopole
energy densities of the classical soliton (LO), the nucleon (NLO) and the $\Delta$ (NLO)
as functions of radius $r$. The values of the three monopole energy densities at $r=0$ are
all found to be $\varepsilon_{0}(0) =\varepsilon^{N,\Delta}_{0}(0) =2.27~\mathrm{GeV}\cdot\mathrm{fm}^{-3}$. The reason is that the value of 
$\delta_{\mathrm{rot}}\varepsilon^{(J)}_0$, which is proportional to the quadrupole energy density $\varepsilon_{2}(r)$ depicted in the right panel of Fig.~\ref{fig:1},
becomes zero at $r=0$. The left panel of Fig.~\ref{fig:1} exhibits a slightly
broader shape of the monopole energy density for the $\Delta$ in comparison with those of 
the nucleon and classical soliton. As explained previously, the rotational corrections
to the monopole energy density are of order $\mathcal{O}(1/N^{2}_{c})$ and hence are
suppressed. The effects of the rotational corrections on the monopole energy densities
can be quantitatively observed by calculating the mass radii
given in Eq.~\eqref{eq:mass_radius},
\begin{align}
\langle r^{2}_{E} \rangle = 0.54 \, \mathrm{fm}^{2} \, (\mathrm{LO}),  \ \ \  \ \
\langle r^{2}_{E} \rangle = 0.57 \, \mathrm{fm}^{2} \, (\mathrm{NLO, \, Nucleon}), \ \ \  \ \
\langle r^{2}_{E} \rangle = 0.64 \, \mathrm{fm}^{2} \, (\mathrm{NLO}, \, \Delta). \ \ \  \ \
\label{eq:rE}
\end{align}
As given in Eq.~\eqref{eq:rE}, the value of the mass radius of the $\Delta$ is larger than those of the classical soliton and nucleon. It reflects the fact that the monopole energy density of the $\Delta$ has a broader shape in comparison with those of the classical soliton and nucleon. In the case of the finite pion mass, the NLO mass radii of the nucleon and the $\Delta$ are found to be finite, since the energy densities are exponentially suppressed in the large distance. However, in the chiral limit, those mass radii diverge because the rotational corrections to the energy densities are proportional to $1/r^{4}$ in the large-$r$, i.e., $\langle r^{2}_{E} \rangle \propto  \int^{\infty}_{0} dr \ r^{4} \varepsilon^{N,\Delta}_{0}(r)$ as given in Eqs.~\eqref{eq:quad_density} and~\eqref{eq:chiral_prop}, which is similar to the isovector charge radius of the nucleon in Ref.~\cite{Adkins:1983ya}.
In the right panel of Fig.~\ref{fig:1}, the numerical result for the quadrupole energy density $\varepsilon_{2}(r)$ of the $\Delta$ is drawn as a function of radius $r$. The quantity $\varepsilon_{2}(r)$ provides  information on how the energy density is deformed from the spherically symmetric shape. It has a peak at around $r=0.4$~fm with a negative sign and its strength is very small compared with $\varepsilon_{0}(r)$. By integrating $\varepsilon_{2}(r)$ over $r$, we can estimate the typical size of $\varepsilon_{2}(r)$ as 
\begin{align}
\int d^{3}r \, \varepsilon_{2}(r)=-\frac{1}{2}\int d^{3}r[\varepsilon^{\Delta}_{0}(r)-\varepsilon^{N}_{0}(r)] =-147~\mathrm{MeV}.
\end{align}
We confirm the relation numerically, first obtained in the large $N_{c}$ limit~\cite{Panteleeva:2020ejw}. The value of the integration of $\varepsilon_{2}(r)$ over $r$ is approximately $10\% \sim \mathcal{O}(1/N^{2}_{c})$ of that of $\varepsilon_{0}(r)$. 
Another interesting property is the mass quadrupole moment given in Eq.~\eqref{eq:mass_moment}. Its value exhibits how the energy density is deformed from the spherically symmetric shape quantitatively and is found to be
\begin{align}
\mathcal{Q}^{ij}_{\sigma'\sigma}=\frac{2}{15}Q^{ij}_{\sigma'\sigma}\int d^{3}r \, r^{2} \varepsilon_{2}(r)  =-0.0181~Q^{ij}_{\sigma'\sigma}\mathrm{GeV}\cdot\mathrm{fm^{2}}.
\end{align}
In the chiral limit, the mass quadrupole moment diverges by the same reason as the NLO mass radius, i.e,  $\varepsilon_{2}(r) \propto \frac{1}{r^{4}}$.

\begin{figure}[htp]
\centering
\includegraphics[scale=0.63]{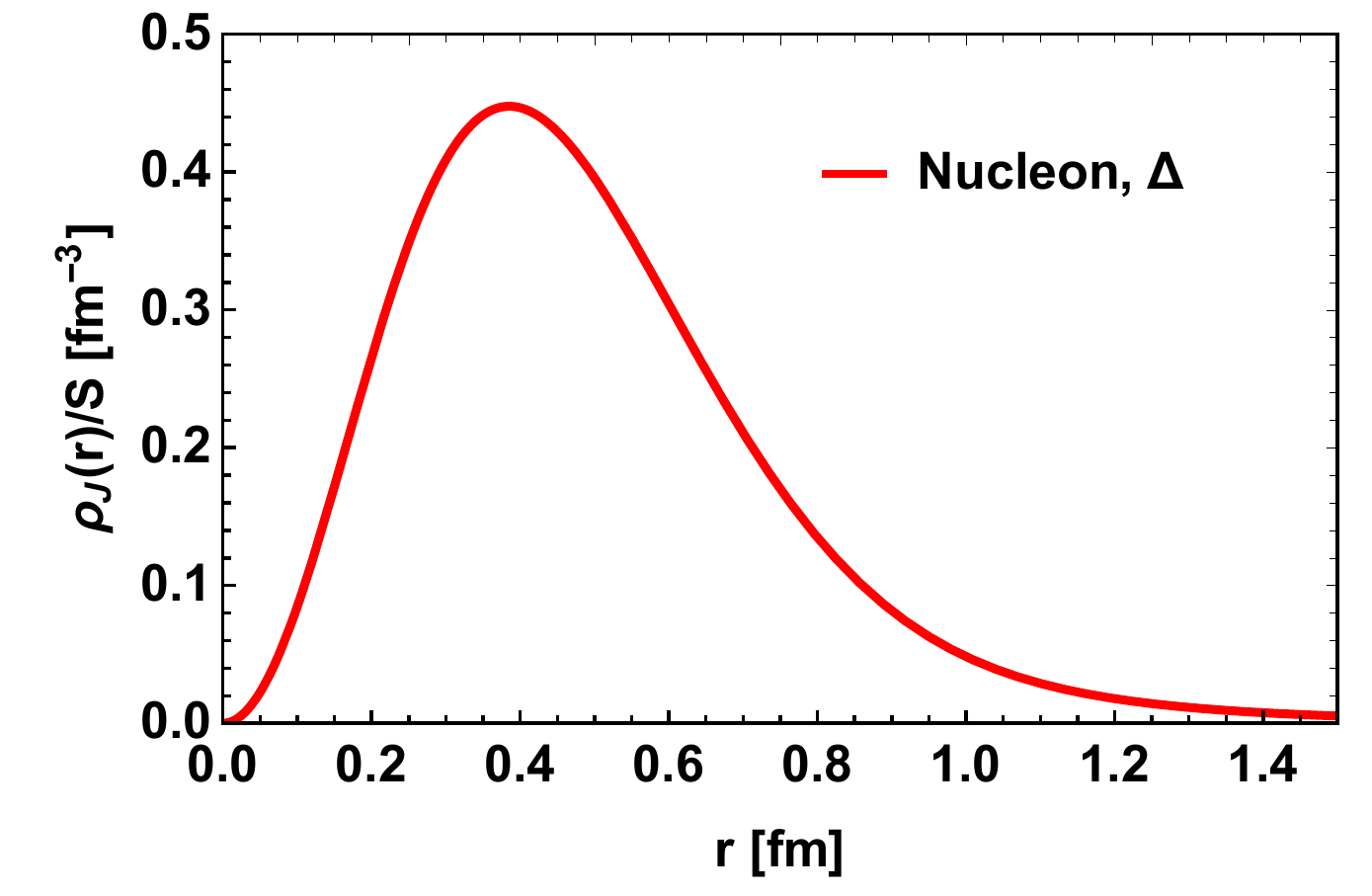}
\caption{The angular momentum densities of the nucleon and the $\Delta$ as functions of $r$ normalized by corresponding baryon spin $S$.}
\label{fig:2}
\end{figure}
Figure~\ref{fig:2} shows the result for the averaged angular momentum density $\rho_{J}(r)$ normalized by the corresponding baryon spin $S$. As shown in  Eqs.~\eqref{eq:angular} and~\eqref{eq:a_den}, $\rho_{J}(r)$ is related to the $0k$-components of the EMT. The integration of $\rho_{J}(r)$ over $r$ yields the constraints on the form factors $\mathcal{J}_{1}(0)$ and $F_{4,0}(0)$ as 
\begin{align}
 \frac{1}{S}\int d^{3}r \, \rho_{J}(r) = 2\mathcal{J}_{1}(0) = \frac{2}{3}F_{4,0}(0) = 1.
 \end{align}
We obtain the numerical results for the mean square radius $\langle r^{2}_{J} \rangle$ of the nucleon and the $\Delta$ as
\begin{align}
\langle r^{2}_{J} \rangle_{N,\Delta}=\frac{\int d^{3}r \, r^{2} \rho_{J}(r)}{\int d^{3}r \, \rho_{J}(r)} = 0.92 \, \mathrm{fm}^{2}.
\end{align}
The results $\langle r^{2}_{J} \rangle$ for the nucleon and the $\Delta$ are degenerate. Since in Eq.~\eqref{eq:rot_density} higher-order corrections to the EMT densities $\delta_{\mathrm{rot}}T^{0k}(\bm{r},\sigma',\sigma)$ are abscent, the averaged angular momentum density of the $\Delta$ differs from that of the nucleon by factor three, and its mean square radius of the $\Delta$ is the same as that of the nucleon.\footnote{Note that in the chiral limit the averaged angular momentum density $\rho_{J}(r)$ decrease as $r^{-4}$ in large distance, so the radius diverges.} As mentioned in the previous Section, the  higher-order corrections to the averaged angular momentum $\delta_{\mathrm{rot}}\rho^{(J)}_{J}(r)$ arise from the higher derivative terms that generate $\Omega^{3}\sim \mathcal{O}(1/N_{c}^{3})$. Since there are no such terms in the present Skyrme Lagrangian, the corresponding corrections $\delta_{\mathrm{rot}}\rho^{(J)}_{J}(r)$ do not appear. Viewed in the argument of the large-$N_{c}$ expansion, the corrections $\delta_{\mathrm{rot}}\rho^{(J)}_{J}(r)$ are strongly suppressed and must be very tiny.

\begin{figure}[htp]
\centering
\includegraphics[scale=0.63]{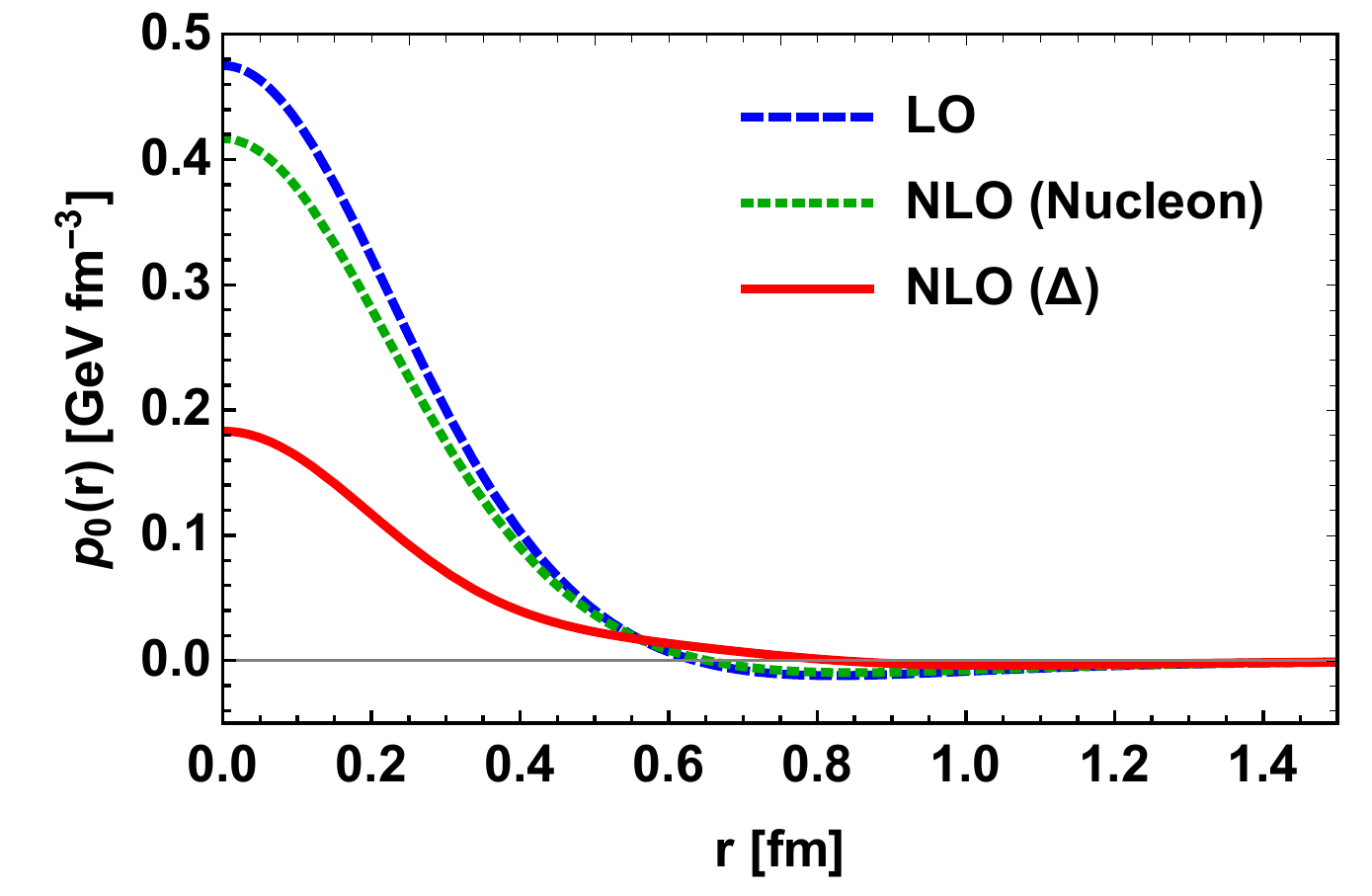}
\includegraphics[scale=0.63]{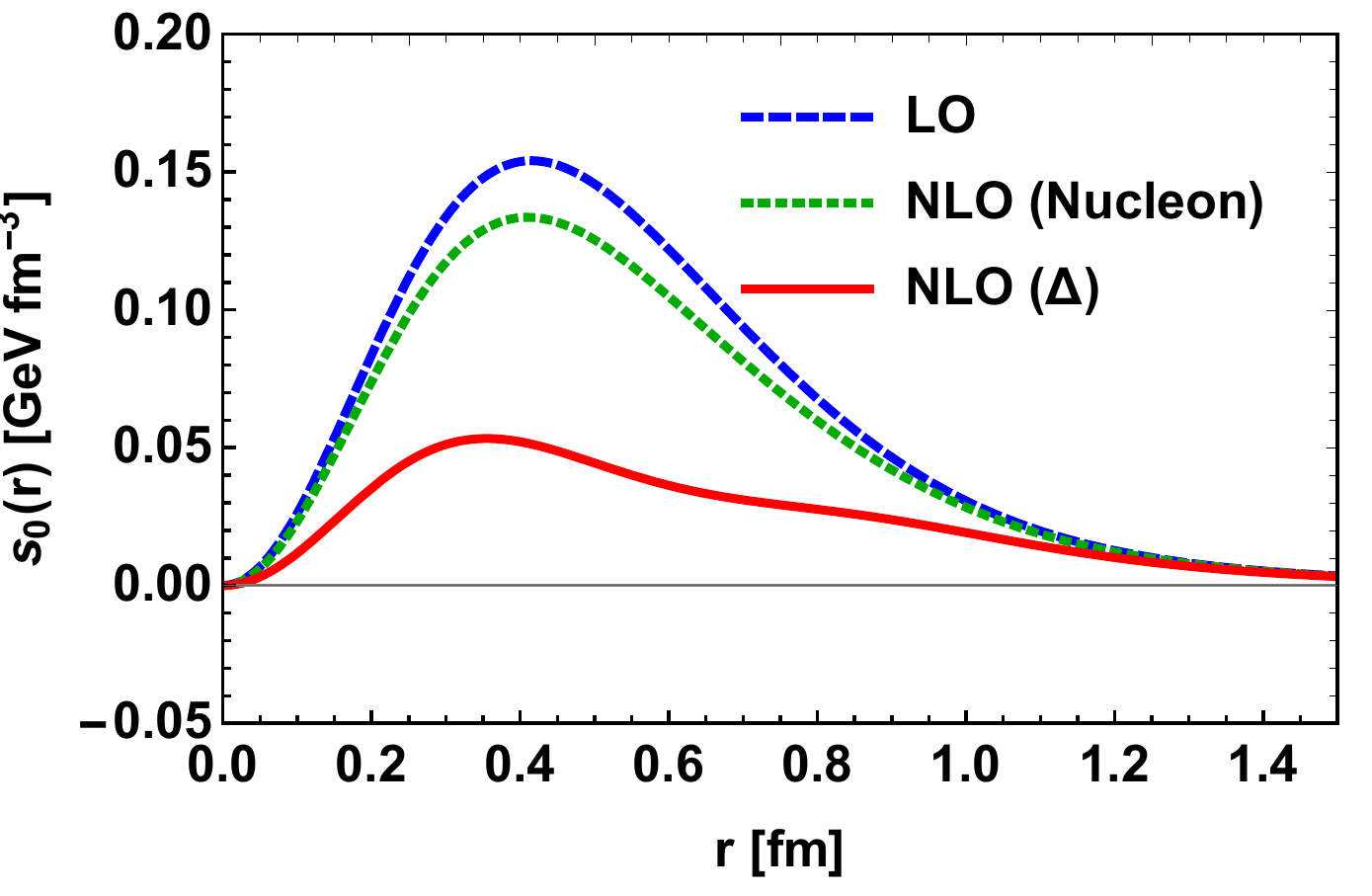}
\caption{The left (right) panel depicts the results for the pressures (shear forces) of the classical soliton, the nucleon, and the $\Delta$ as functions of radius $r$. The dashed, dotted and solid curves draw the classical soliton, the nucleon, and the $\Delta$, respectively. Note that in the case of  results of the NLO pressures they are reconstructed from Eq.~\eqref{eq:EMT_differential} by using the approximated shear forces to comply with the von Laue condition.}
\label{fig:3}
\end{figure}
The left panel of Fig.~\ref{fig:3} shows the results of the pressure densities for the classical soliton (LO), the nucleon (NLO), and the $\Delta$ (NLO) as functions of radius $r$. The LO pressure naturally complies with the von Laue condition that is equivalent to the equation of motion. However, as discussed in the previous Section the NLO pressures do not satisfy the stability condition. Thus, we adopt a strategy that preserves the chiral symmetry and satisfies the stability condition. We first take the variation of the soliton mass, and then quantize the soliton. By treating the rotational corrections as a small perturbation, we then obtain the approximated shear force density $s_{0}(r)$. With using $s_{0}(r)$, we reconstruct the pressure $p_{0}(r)|_{\mathrm{reconst}}$ from Eq.~\eqref{eq:EMT_differential}, so that the pressure density meets the stability condition. Thus, the LO and NLO pressures in Fig.~\ref{fig:3} satisfy the stability condition.  In the meantime, satisfying the stability condition implies that the pressure density has at least one nodal point ($r_{0}$) where the pressure density vanishes. For the inner part of $r_{0}$~($p_{0}>0$) repulsive force dominates, whereas for the outer part of $r_{0}$~($p_{0}<0$) attractive force governs. Thus, we obtain the nodal points $r_{0}$ for the classical soliton, nucleon and $\Delta$ numerically as follows:
\begin{align}
r_{0} =0.64 \, \mathrm{fm} \, (\mathrm{LO}),  \ \ \  \ \
r_{0} = 0.65 \, \mathrm{fm} \, (\mathrm{NLO, \, Nucleon}), \ \ \  \ \
r_{0} = 0.83 \, \mathrm{fm} \, (\mathrm{NLO}, \, \Delta). \ \ \  \ \
\label{eq:r0}
\end{align}
While the value of $r_{0}$ for the nucleon and the classical soliton are comparable, that for the $\Delta$ is found to be much larger than the results for the nucleon and the classical soliton. It indicates that the $\Delta$ is mechanically spreading more widely compared with them. The right panel of Fig.~\ref{fig:3} illustrates the results of the approximated shear force densities for the $\Delta$, the nucleon and the classical soliton. To comply with the stability condition the $D$-term $D_{0}(0)$ should be negative, which means that the integral (weighted by $r^2$) of the shear force densities over all values of $r$ should be positive. (see Eq.~\eqref{eq:D_term}). Indeed, the shear force densities for them are always positive. While the $s_{0}(r)$ for the $\Delta$ has a wide-spreading shape in comparison with those for the classical soliton and nucleon, the magnitude of $s_{0}(r)$ for the $\Delta$ is considerably small compared with those for the others. 

\begin{figure}[htp]
\centering
\includegraphics[scale=0.63]{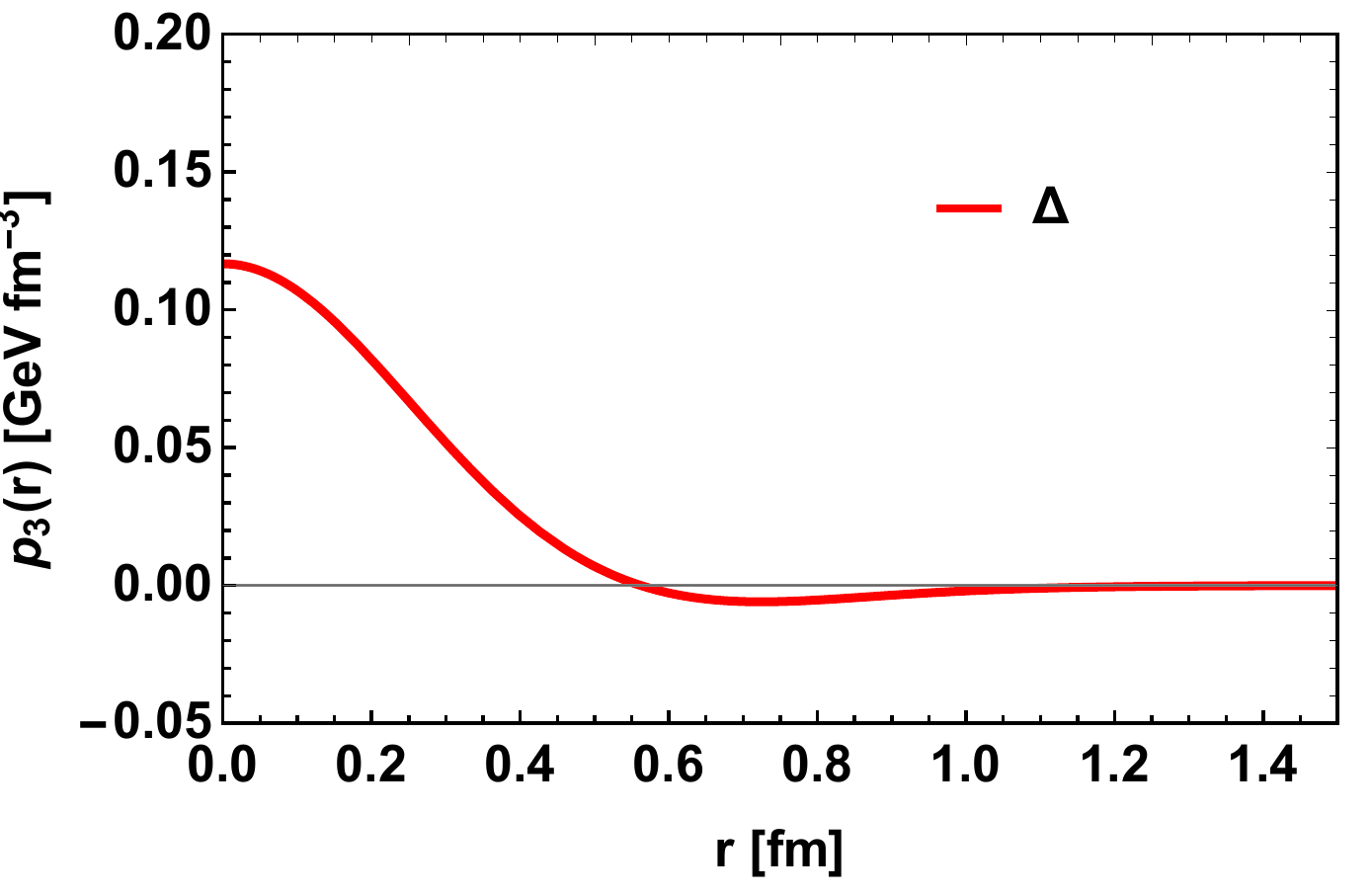}
\includegraphics[scale=0.63]{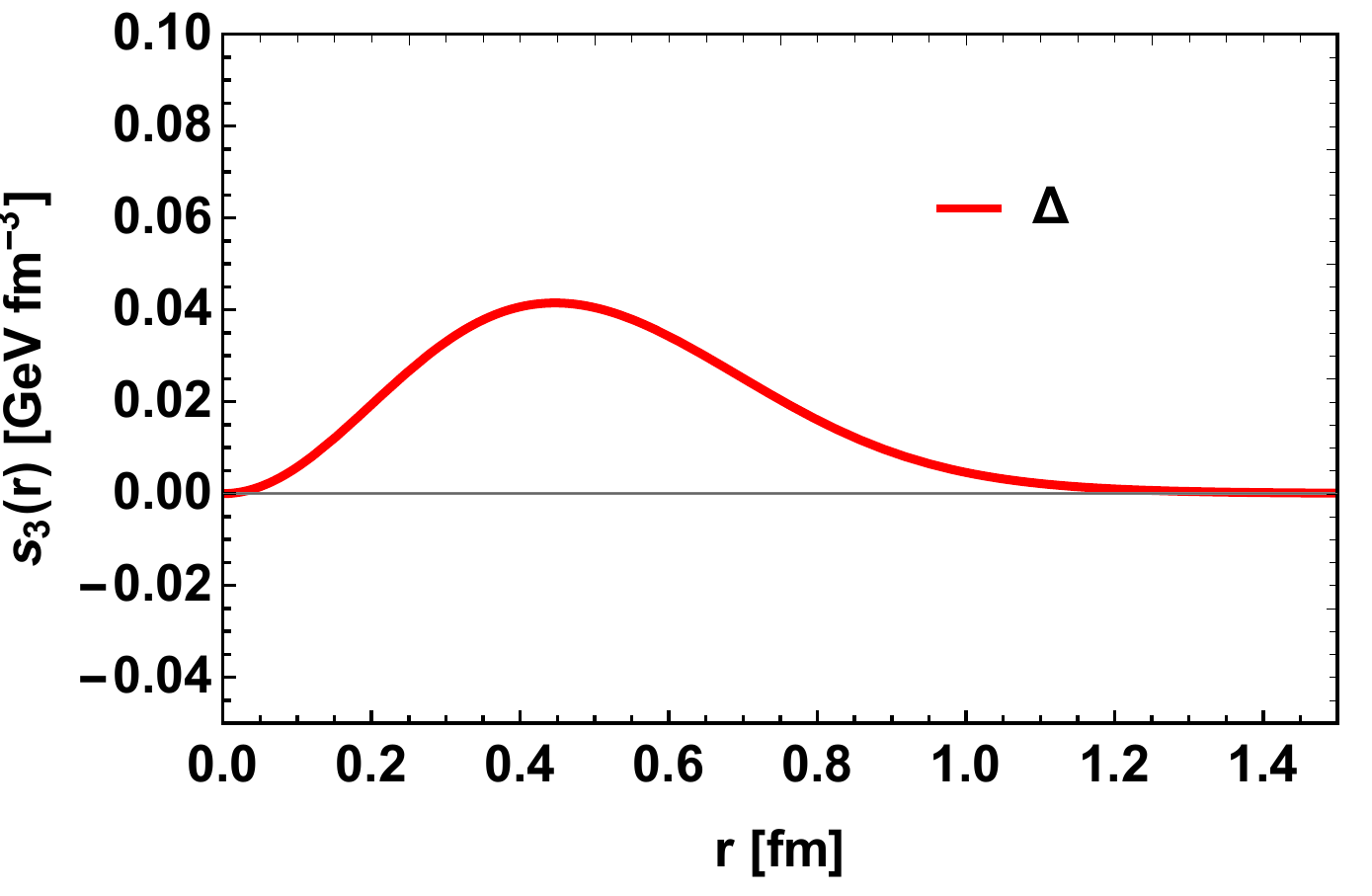}
\caption{The left (right) panel depicts the results for the quadrupole pressures (shear forces) of the $\Delta$ as a function of radius $r$. The quadrupole pressure density $p_{3}(r)$  is reconstructed from Eq.~\eqref{eq:EMT_differential} by using the approximated quadrupole shear force density $s_{3}(r)$ in order to comply with the von Laue condition.}
\label{fig:4}
\end{figure}
Figure~\ref{fig:4} shows the result of the quadrupole pressure (shear force) density for the $\Delta$. The quadrupole pressure $p_{3}(r)|_{\mathrm{reconst}}$ is reconstructed, as the NLO pressures $p^{N,\Delta}_{0}(r)|_{\mathrm{reconst}}$ were derived from Eq.~\eqref{eq:EMT_differential}. Thus, $p_{3}(r)|_{\mathrm{reconst}}$ also complies with the stability condition given in Eq.~\eqref{eq:von_Laue}, and has a nodal point located at $r_{0} =0.56 \, \mathrm{fm}$. The shapes of $p_{3}(r)|_{\mathrm{reconst}}$ and $s_{3}(r)$ are similar to those of $p_{0}(r)$ and $s_{0}(r)$, respectively. 

\begin{figure}
\centering
\includegraphics[scale=0.63]{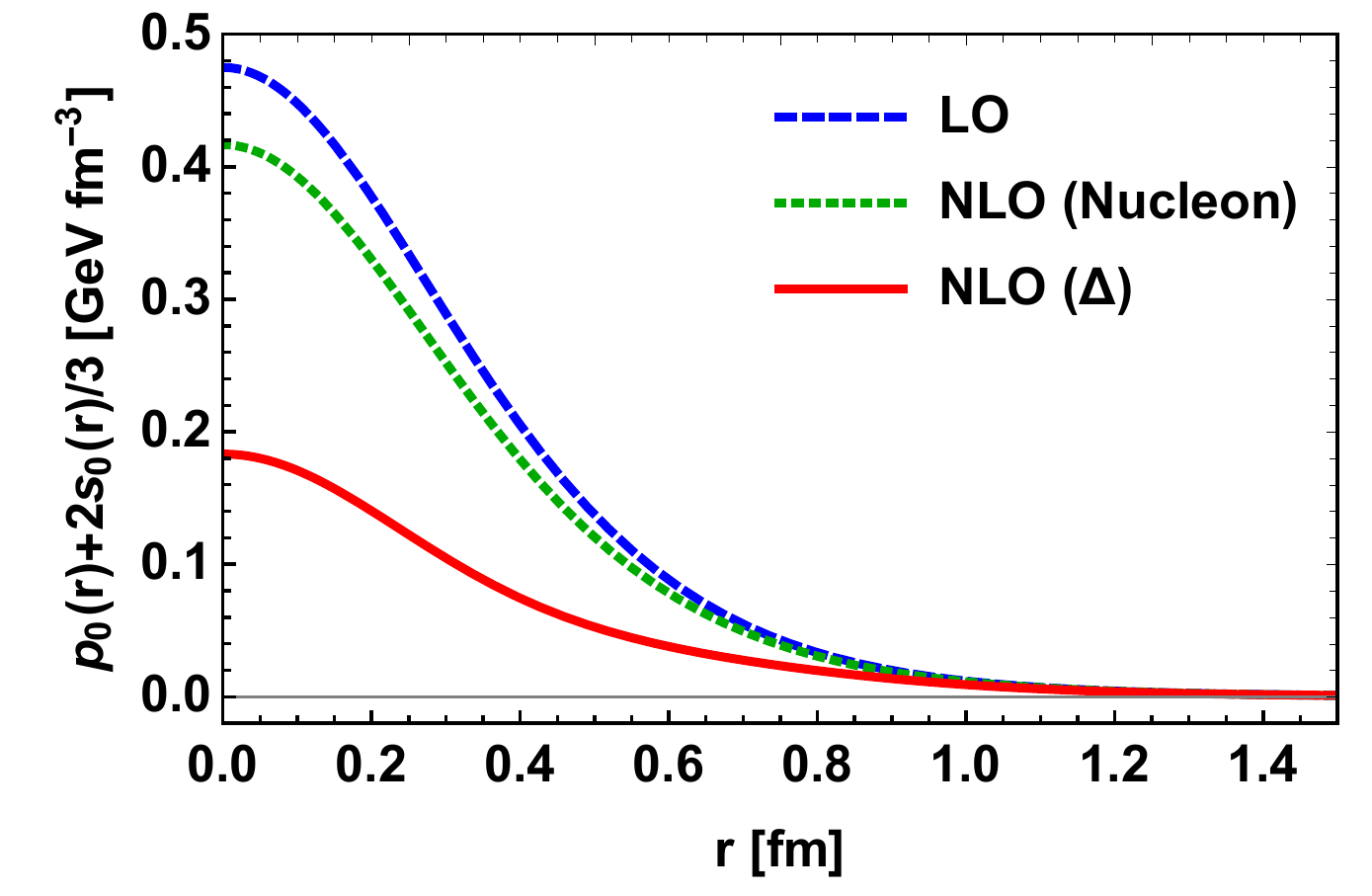}
\includegraphics[scale=0.63]{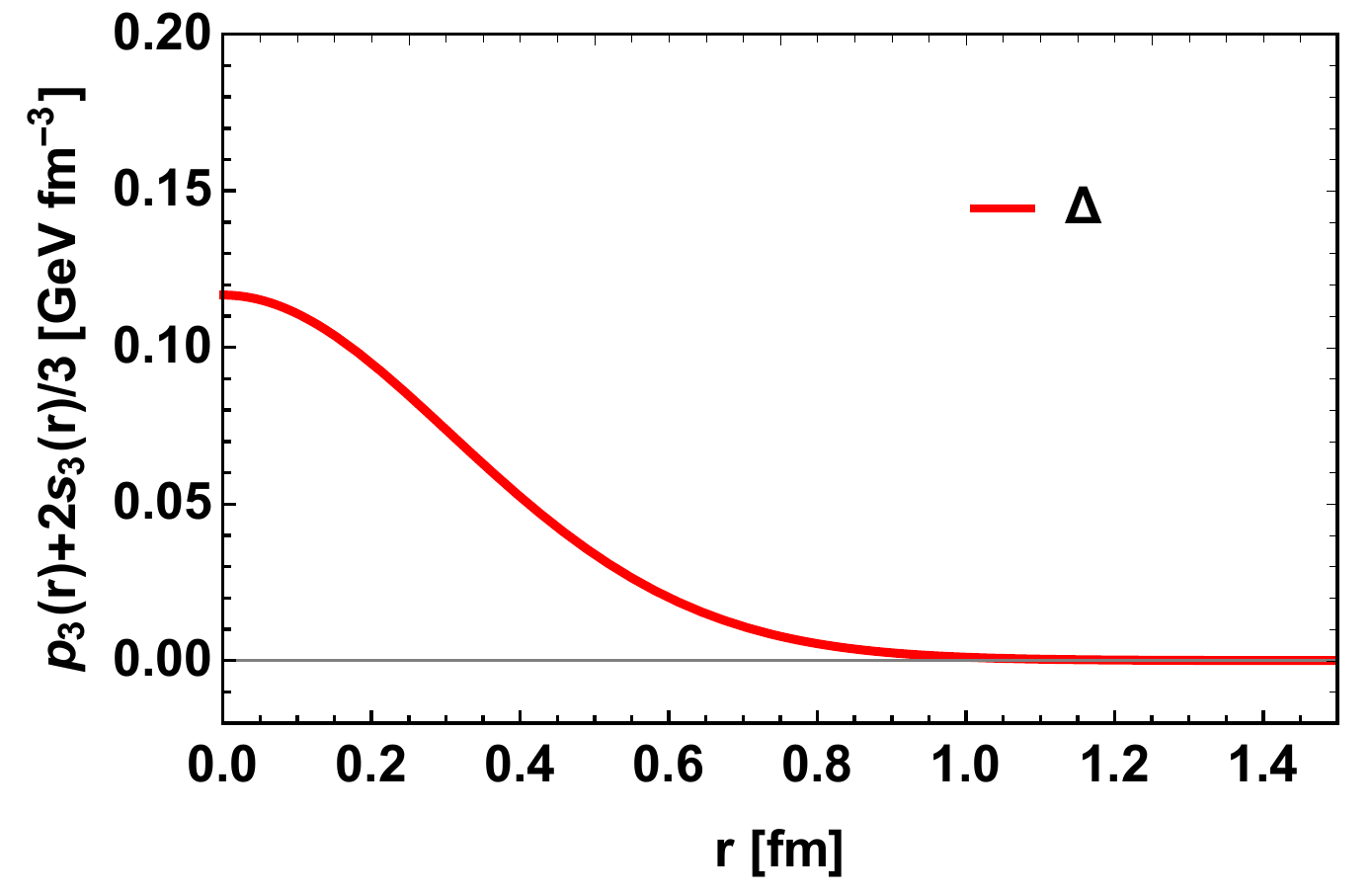}
\caption{The left panel depicts the results for the monopole normal forces $p_{0}(r)|_{\mathrm{reconst}}+\frac{2}{3}s_{0}(r)$, given in Eq.~\eqref{eq:force}, acting on the radial area element, and simultaneously shows local stability condition in Eq.~\eqref{eq:local} for the classical soliton, nucleon and unpolarized $\Delta$. The right panel shows the result for the quadrupole normal force $p_{3}(r)|_{\mathrm{reconst}}+\frac{2}{3}s_{3}(r)$. The NLO and quadrupole pressure densities are reconstructed from Eq.~\eqref{eq:EMT_differential} by using the approximated shear force densities $s_{0}(r)$ and $s_{3}(r)$ in order to comply with the von Laue condition.}
\label{fig:5}
\end{figure}
The left panel of Fig.~\ref{fig:5} presents the results of the monopole normal force components $p_{0}(r)|_{\mathrm{reconst}}+\frac{2}{3}s_{0}(r)$ acting on the radial area element for the classical soliton, the nucleon and the unpolarized $\Delta$. When it comes to a spherically symmetric baryon, the monopole normal force can be directly related to the local stability condition~\eqref{eq:local} stating that the force is always directed outwards.
As for the polarized $\Delta$, the contributions of the quadrupole densities to the normal force come into play. The result of the quadrupole normal force $p_{3}(r)|_{\mathrm{reconst}}+\frac{2}{3}s_{3}(r)$ is shown in the right panel of Fig.~\ref{fig:5}, except for that of the nullified  densities $s_{2}(r)$ and $p_{2}(r)$. It is found that the monopole and quadrupole normal forces are positive over the whole region of $r$. For the large distance in the chiral limit, the monopole normal force keeps the positivity, which means that the local stability condition is satisfied. Note that the quadrupole normal force has a positive sign
\begin{align}
0<p^{N,\Delta}_{0}(r)\bigg{|}_{\mathrm{reconst}}+\frac{2}{3}s^{N,\Delta}_{0}(r)= F^{2}_{\pi}\frac{R^{4}_{0}}{r^{6}}  \cdots, \ \ \ \ \ 0<p_{3}(r)\bigg{|}_{\mathrm{reconst}}+\frac{2}{3}s_{3}(r) =  \frac{56}{5}\frac{1}{I^{2}e^{2}}\frac{R^{8}_{0}}{r^{10}} \cdots.
\end{align}
However, we do not know how the quadrupole normal force is related to the local stability conditions so far. Another interesting quantity is the mechanical radius. As defined in Ref.~\cite{Polyakov:2018zvc}, the mechanical radius is obtained by
\begin{align}
\langle r_{0}^{2} \rangle_{\mathrm{mech}} = 0.61 \, \mathrm{fm}^{2} \, (\mathrm{LO}),  \ \ \  \ \
\langle r_{0}^{2} \rangle_{\mathrm{mech}} = 0.63 \, \mathrm{fm}^{2} \, (\mathrm{NLO, \, Nucleon}), \ \ \  \ \
\langle r_{0}^{2} \rangle_{\mathrm{mech}} = 0.85 \, \mathrm{fm}^{2} \, (\mathrm{NLO}, \, \Delta). \ \ \  \ \
\end{align}
Similar to Eqs.~\eqref{eq:rE} and~\eqref{eq:r0} the value of $\langle r_{0}^{2} \rangle_{\mathrm{mech}}$ for the $\Delta$ is larger than those for the classical soliton and nucleon, which verifies that indeed $\Delta$ is mechanically spreading more widely compared with those for the classical soliton and the nucleon.
Furthermore, we define the quadrupole mechanical radius $\langle r_{3}^{2} \rangle_{\mathrm{mech}}$ in Eq.~\eqref{eq:mech}, and its numerical value is found to be
\begin{align}
\langle r_{3}^{2} \rangle_{\mathrm{mech}} = 0.33 \, \mathrm{fm}^{2}. \,
\end{align}
The quadrupole normal force is very compact in comparision with monopole one.

\begin{figure}[htp]
\centering
\includegraphics[scale=0.629]{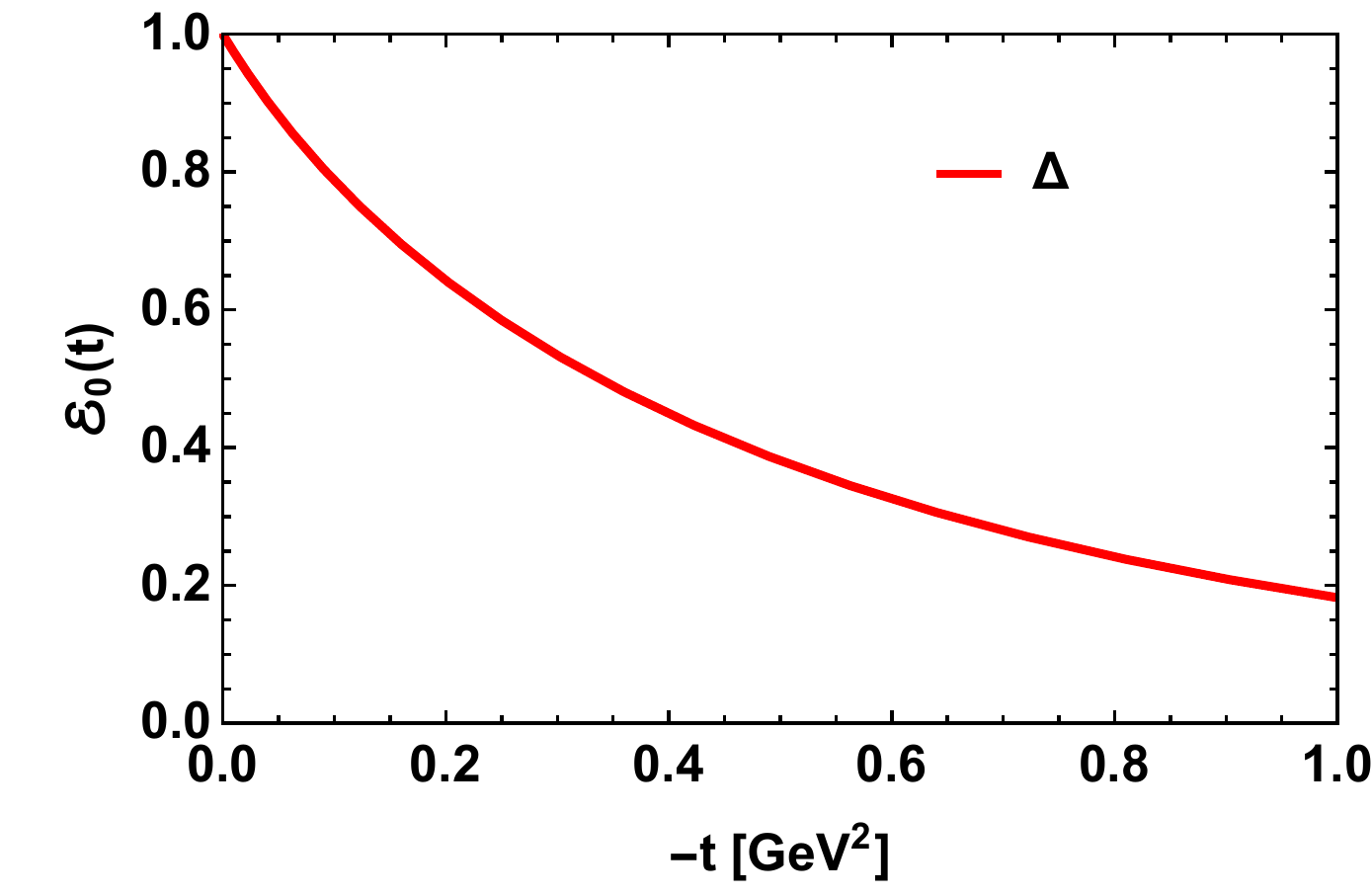}
\includegraphics[scale=0.629]{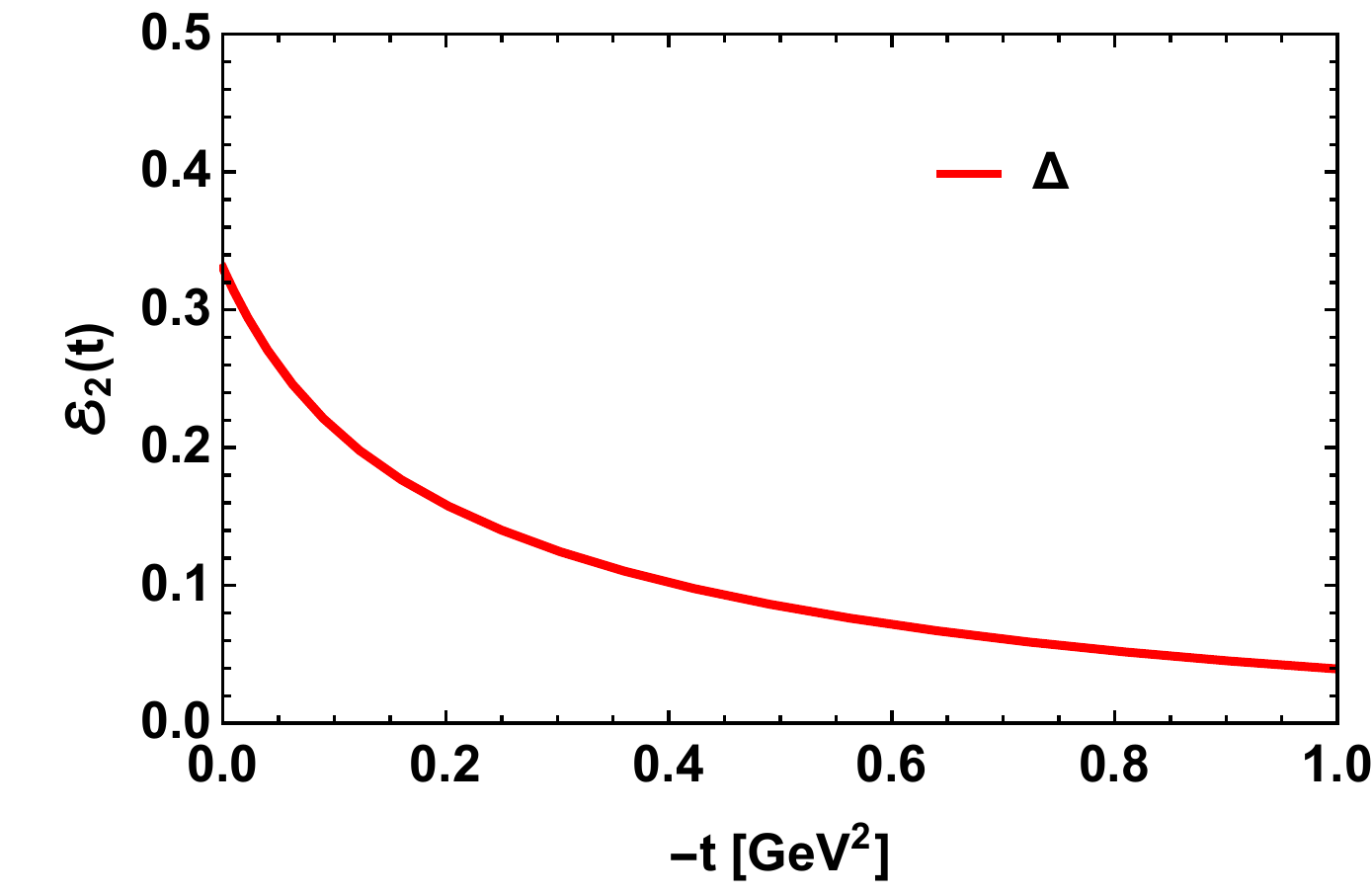}
\includegraphics[scale=0.629]{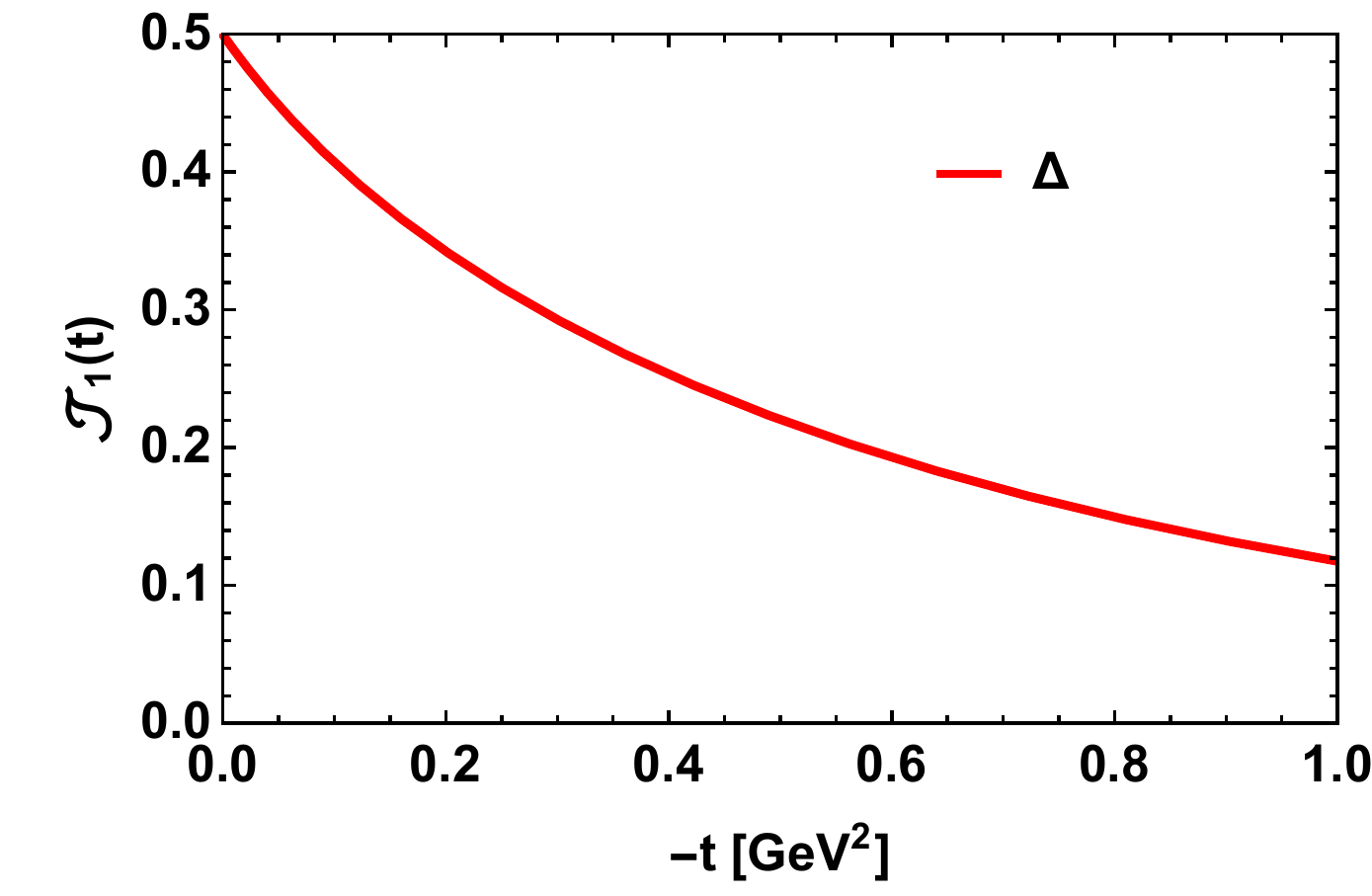}
\includegraphics[scale=0.629]{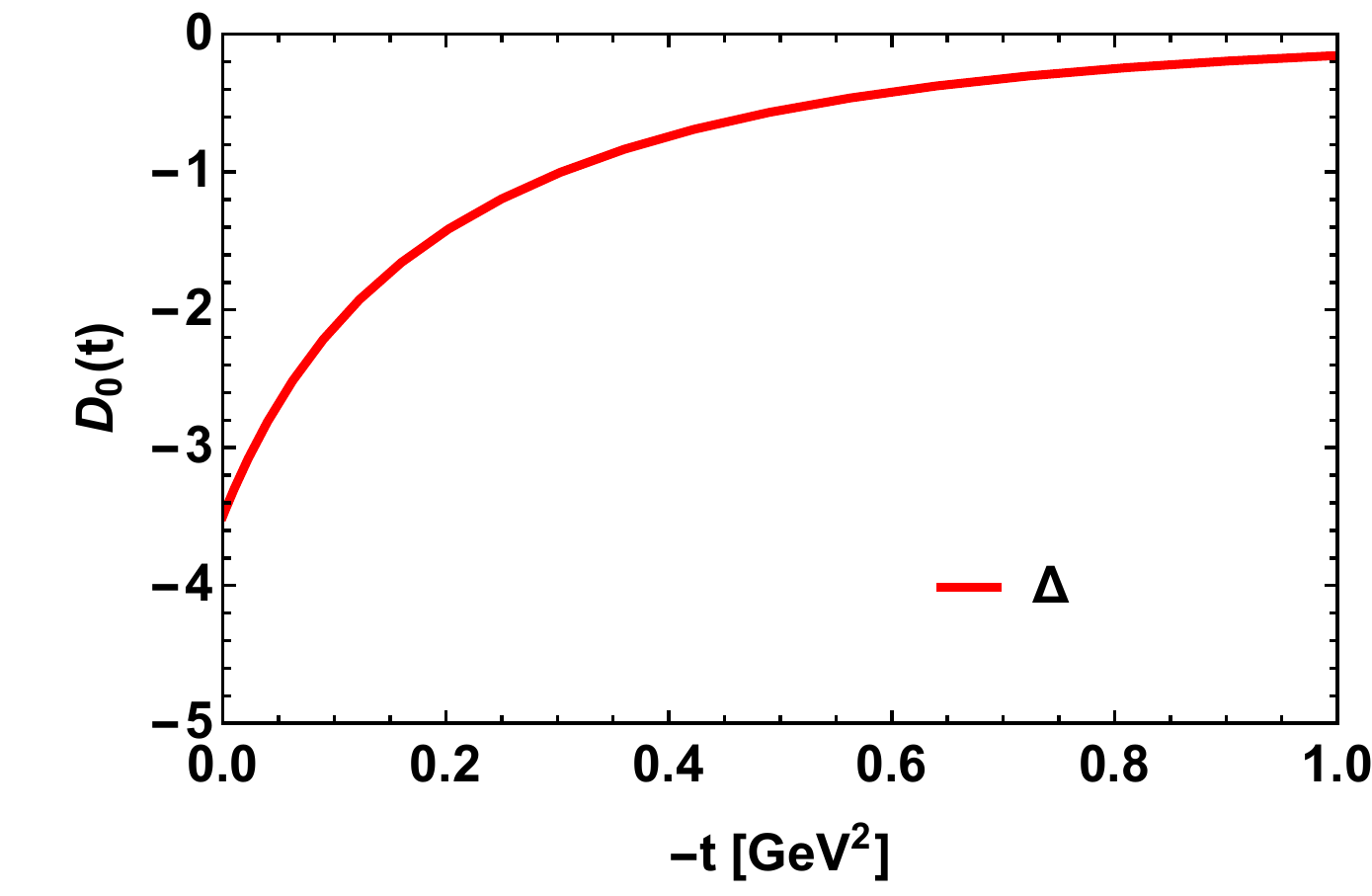}
\includegraphics[scale=0.629]{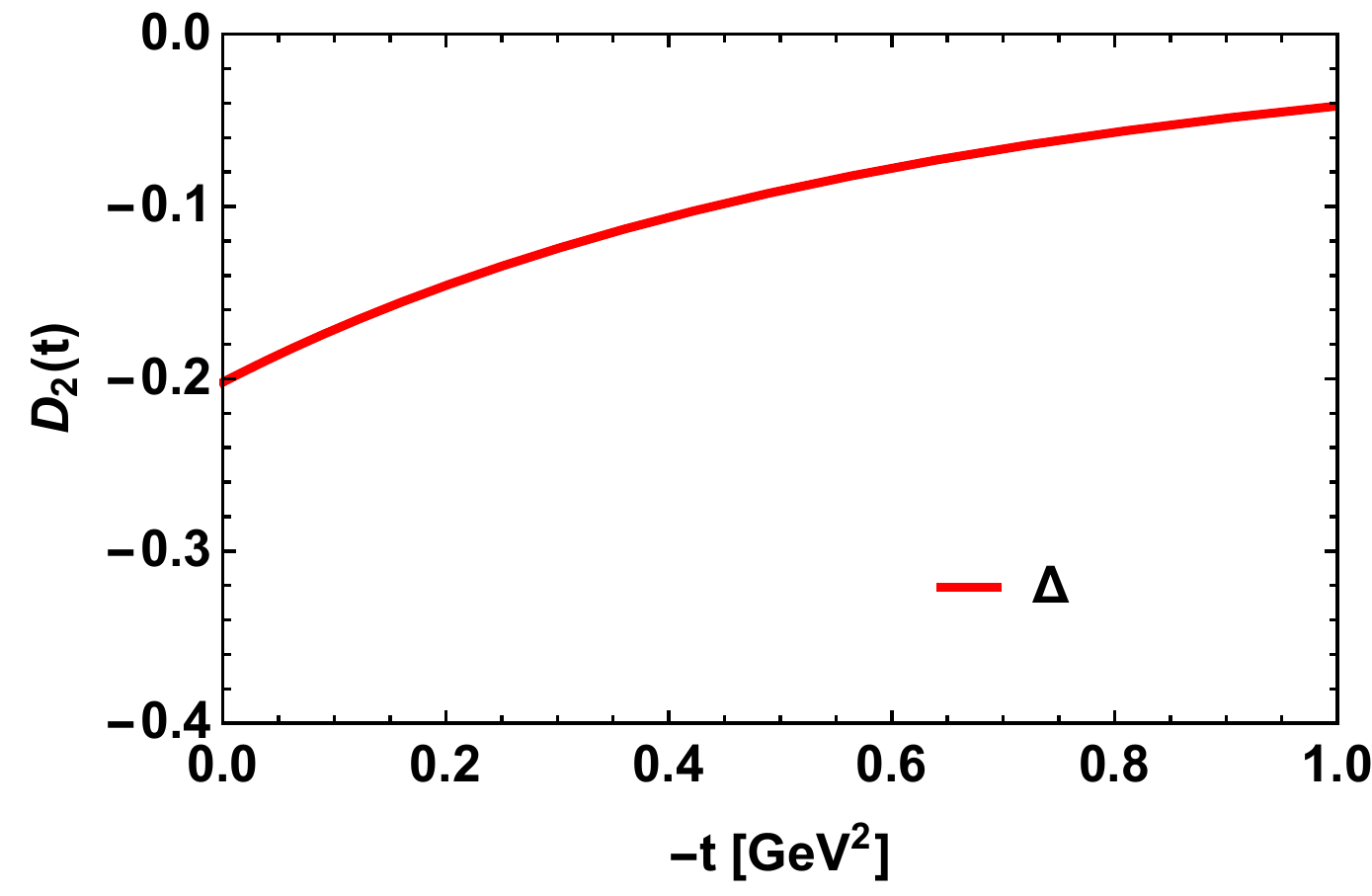}
\includegraphics[scale=0.629]{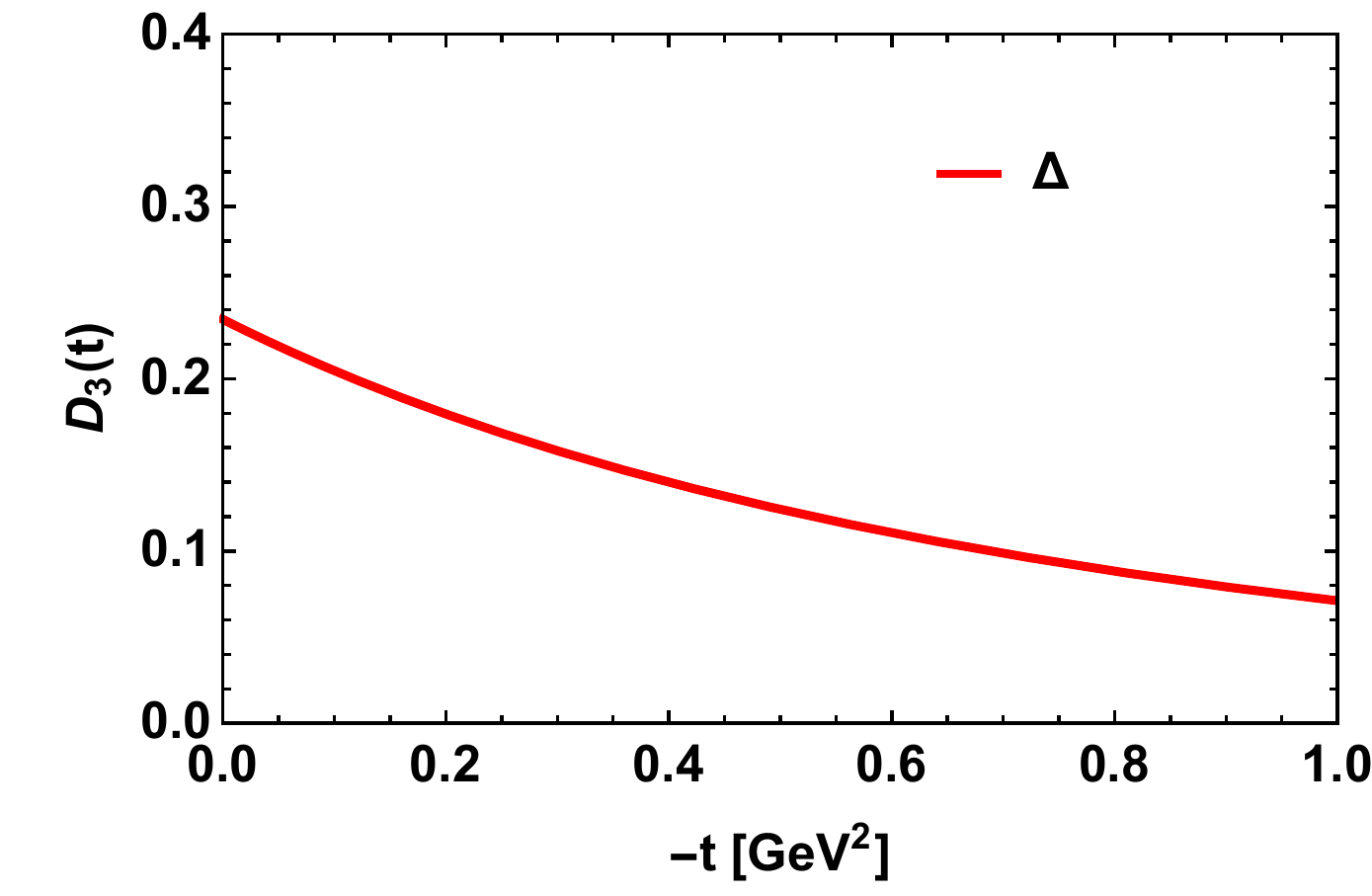}
\caption{Gravitational multipole form factors of the $\Delta$ as functions of squared momentum transfer $t$.}
\label{fig:6}
\end{figure}
Finally, we discuss the GFFs and GMFFs. As shown in Sec.~\ref{sec:2}, the GMFFs are expressed in terms of the GFFs and related the EMT densities as shown in Eqs.~\eqref{eq:EMTFF_E},~\eqref{eq:EMTFF_J} and~\eqref{eq:EMTFF_D}. In the Skyrme model, the energy $\mathcal{E}_{0}(t)$ and angular momentum $\mathcal{J}_{1}(t)$ form factors satisfy the constraints $\mathcal{E}_{0}(0)=F_{1,0}(0)=1$ and $\mathcal{J}_{1}(0)=\frac{1}{3}F_{4,0}(0)=\frac{1}{2}$. Besides the octupole angular momentum form factor $\mathcal{J}_{3}(t)$ is assume to be zero, i.e., $\mathcal{J}_{3}(0)=-\frac{1}{6}[F_{4,0}(0)+F_{4,1}(0)]=0$, because the corresponding density is suppressed in the large-$N_{c}$ expansion. As a result, we get the relation $F_{4,1}(0)=-F_{4,0}(0)=-3/2$. There is no additional constraint on $\mathcal{E}_{2}(t),D_{0}(t),D_{2}(t)$ and $D_{3}(t)$. Therefore, we determine the moments of the GMFFs from Eqs.~\eqref{eq:EMTFF_E},~\eqref{eq:EMTFF_J} and~\eqref{eq:EMTFF_D} as \footnote{By assuming that the pressure $p_{n}(r)$ and shear force $s_{n}(r)$ densities vanish at large-$r$ faster than any power of $r$, one can arrive at the general relation between them as follows~\cite{Goeke:2007fp}: $\int_{0}^{\infty} dr \, r^{N} s_{n}(r) = -\frac{3(N+1)}{2(N-2)}\int_{0}^{\infty} dr \, r^{N} p_{n}(r)$ for $N>-1$, which arise from the differential equation~\eqref{eq:EMT_differential}. }
\begin{align}
\mathcal{E}_{2}(0) &= -\frac{M_{\Delta}}{15}\int d^{3}r \,r^{2} \varepsilon_{2}(r)=-\frac{1}{6}[F_{1,0}(0)+F_{1,1}(0)-4F_{5,0}(0)]= 0.34, \cr
D_{0}(0) &= M_{\Delta}\int d^{3}r \, r^{2} p^{\Delta}_{0}(r) =-\frac{4}{15}M_{\Delta}\int d^{3}r \, r^{2} s^{\Delta}_{0}(r) =F_{2,0}(0)-\frac{16}{3}F_{5,0}(0)= -3.53 < 0, \cr
D_{2}(0) &=\frac{2}{5}M_{\Delta}\int d^{3}r \, r^{2} p_{3}(r)=-\frac{8}{75}M_{\Delta}\int d^{3}r \, r^{2} s_{3}(r)=\frac{4}{3}F_{5,0}(0)=-0.20, \cr
D_{3}(0) &= -\frac{1}{140}M_{\Delta}^{3}\int d^{3}r \, r^{4} p_{3}(r)= \frac{2}{735}M_{\Delta}^{3}\int d^{3}r \, r^{4} s_{3}(r)=\frac{1}{6}[-F_{2,0}(0)-F_{2,1}(0)+4F_{5,0}(0)]=0.24.
\end{align}
The energy quadrupole form factor is obtained as $\mathcal{E}_{2}(0)=0.34$, and it is related to the mass quadrupole moment given in Eq.\eqref{eq:mass_moment}. The value of the $D$-term\footnote{The discrepancy of $D$-term between this work and Ref.~\cite{Perevalova:2016dln, Panteleeva:2020ejw} arises because of the different masses. They have used the LO mass, whereas we take the NLO masses to keep a consistency in this work.} $D_{0}(0)$ of the $\Delta$ is found to be $-3.53$. The quadrupole form factors $D_{2}(0)$ and $D_{3}(0)$  turn out to be $-0.20$ and $0.24$, respectively. Those quadrupole form factors are related to the generalized $D$-terms given in Eq.~\eqref{eq:D_rel} and determined as 
\begin{align}
{\mathcal{D}}^{\Delta}_{0}= -3.53, \ \  {\mathcal{D}}^{N}_{0}= -3.63, \  \ {\mathcal{D}}_{2}= 0, \ \ {\mathcal{D}}_{3}  = -0.50.
\end{align}
The generalized $D$-terms of the present work are comparable with those of Ref.~\cite{Perevalova:2016dln, Panteleeva:2020ejw}. The numerical results for the $\Delta$ GMFFs and GFFs as functions of $t$ are shown in Fig.~\ref{fig:6} and Fig.~\ref{fig:7}, respectively. We restrict ourselves in the range of $0<(-t)<1~\mathrm{GeV}^{2}$ because of the validity of the large-$N_{c}$ expansion, i.e., $|t|\ll M^{2}_{\Delta}$. We find that typical sizes of the quadrupole form factors $\mathcal{E}_{2}(t)$ $D_{2}(t)$ and $D_{3}(t)$ are relatively small  in comparison with those of the monopole form factors $\mathcal{E}_{0}(t)$ and $D_{0}(t)$.
\begin{figure}[htp]
\centering
\includegraphics[scale=0.629]{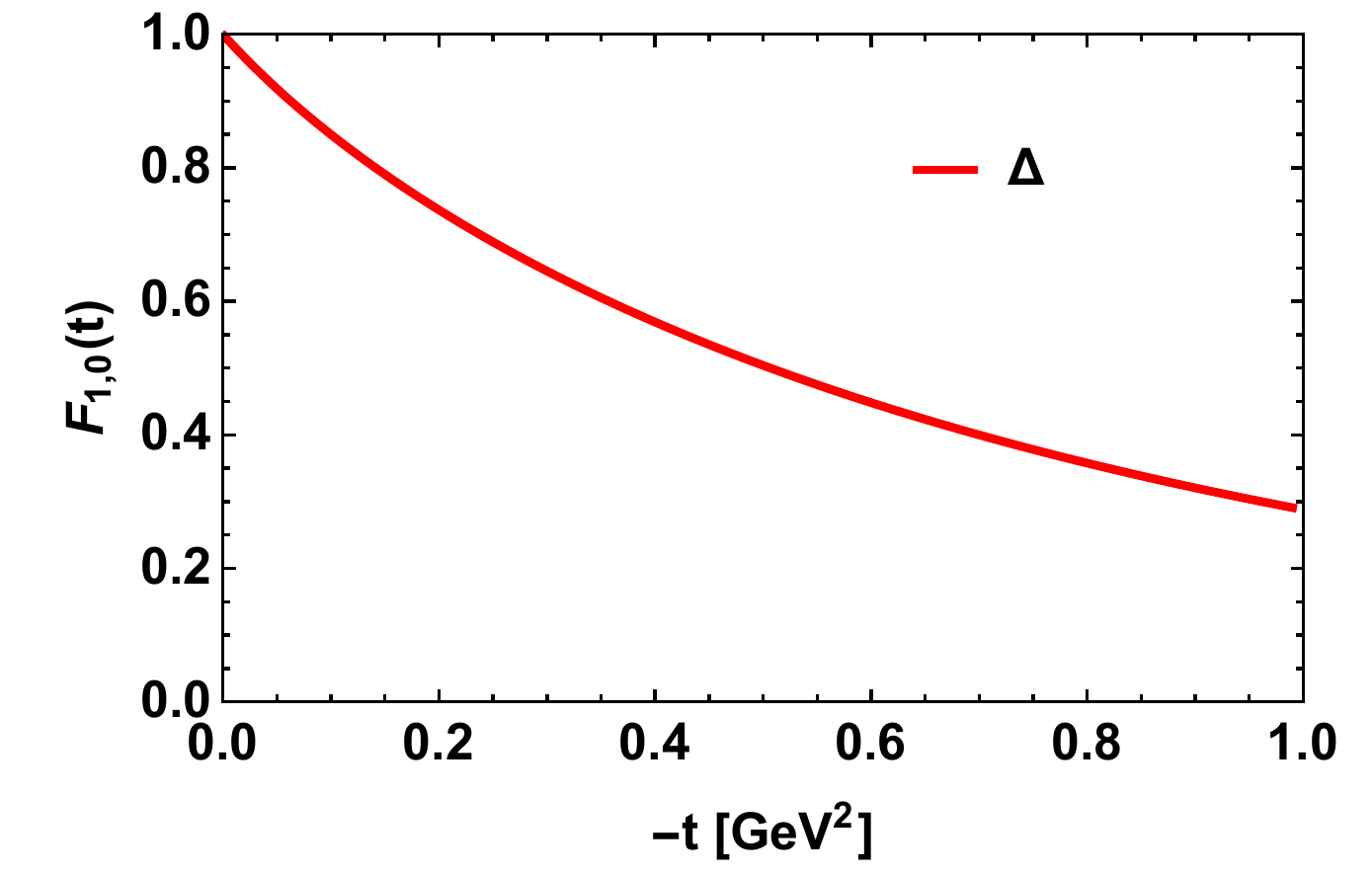}
\includegraphics[scale=0.629]{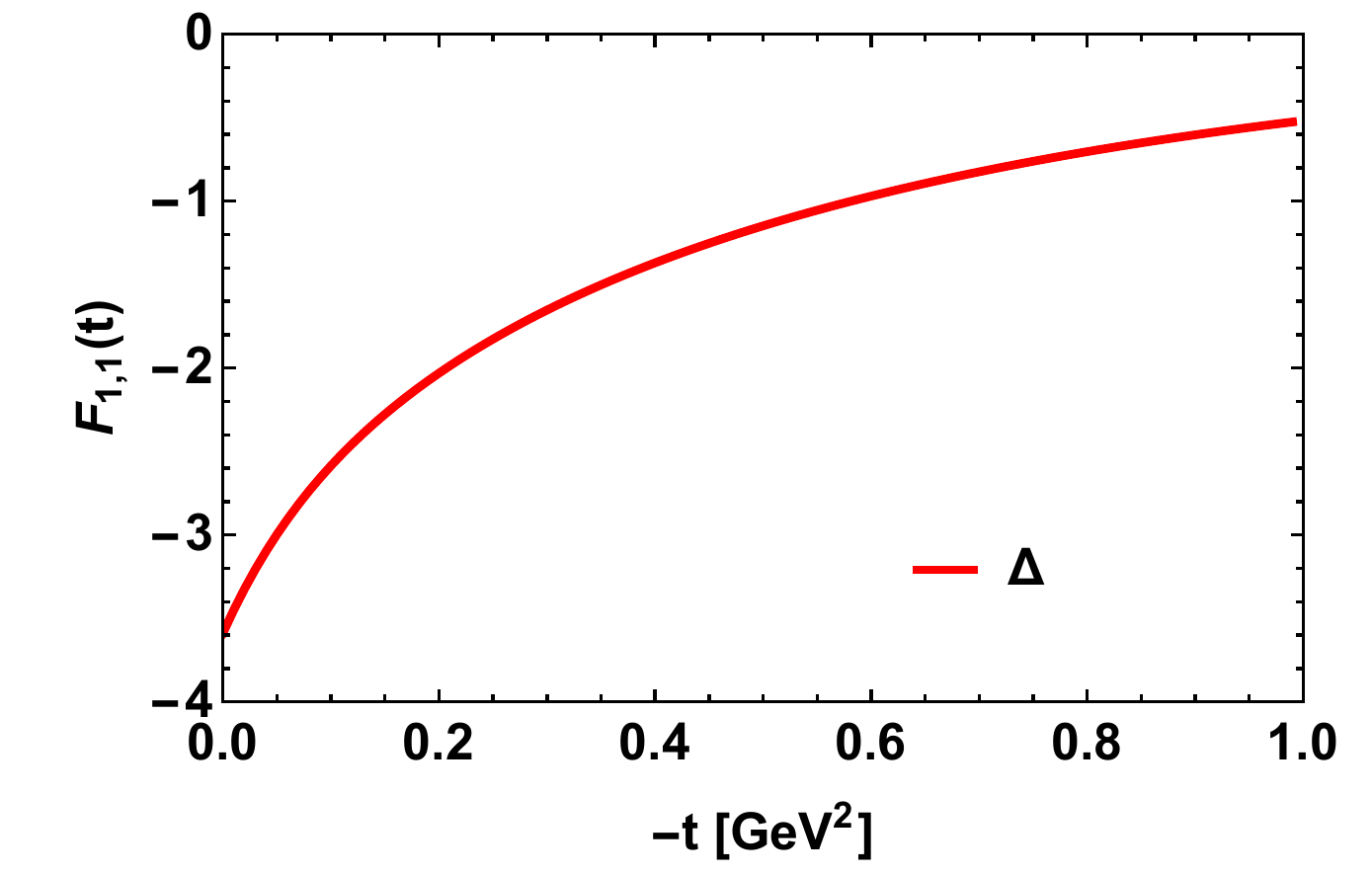}
\includegraphics[scale=0.629]{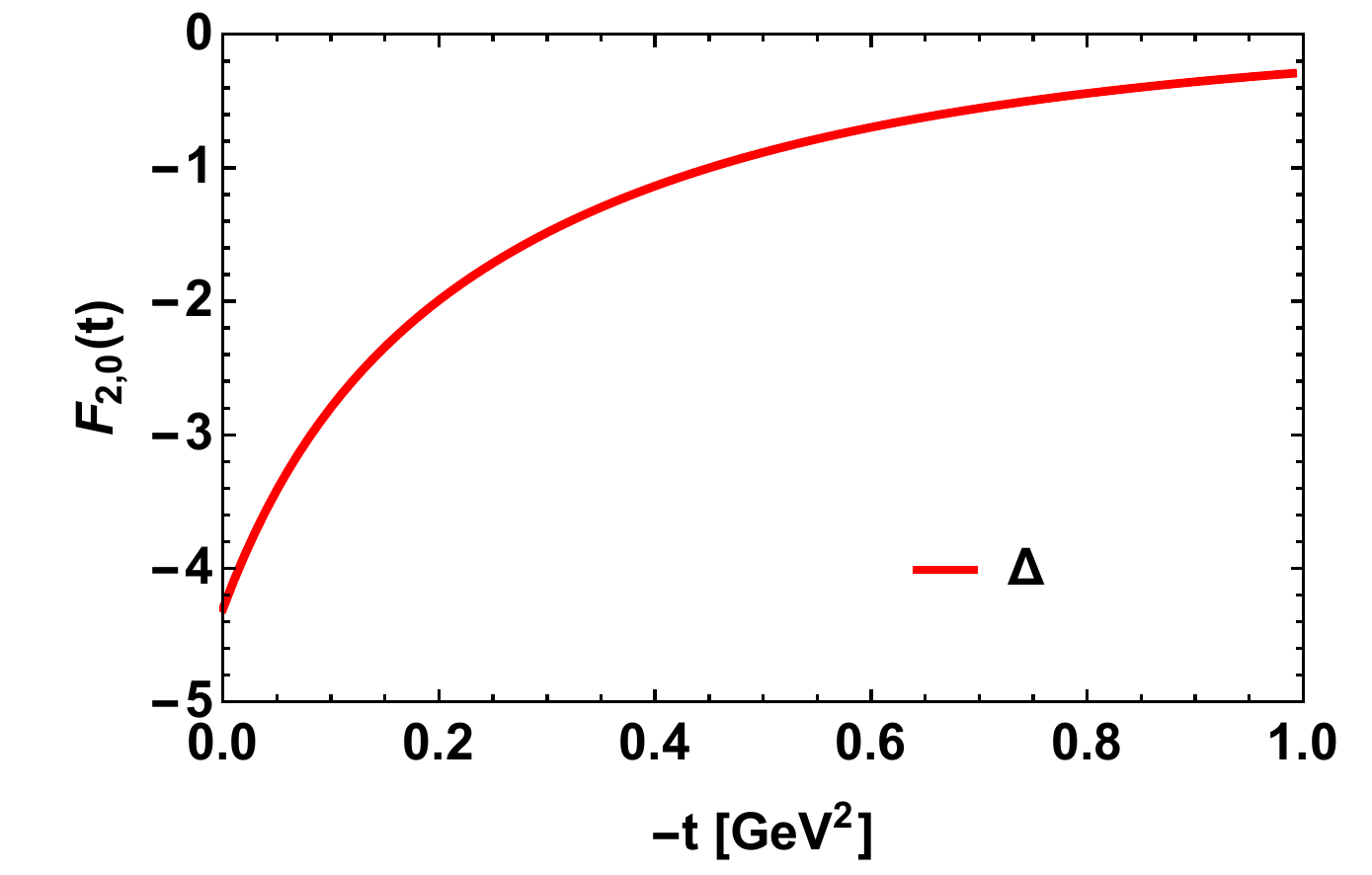}
\includegraphics[scale=0.629]{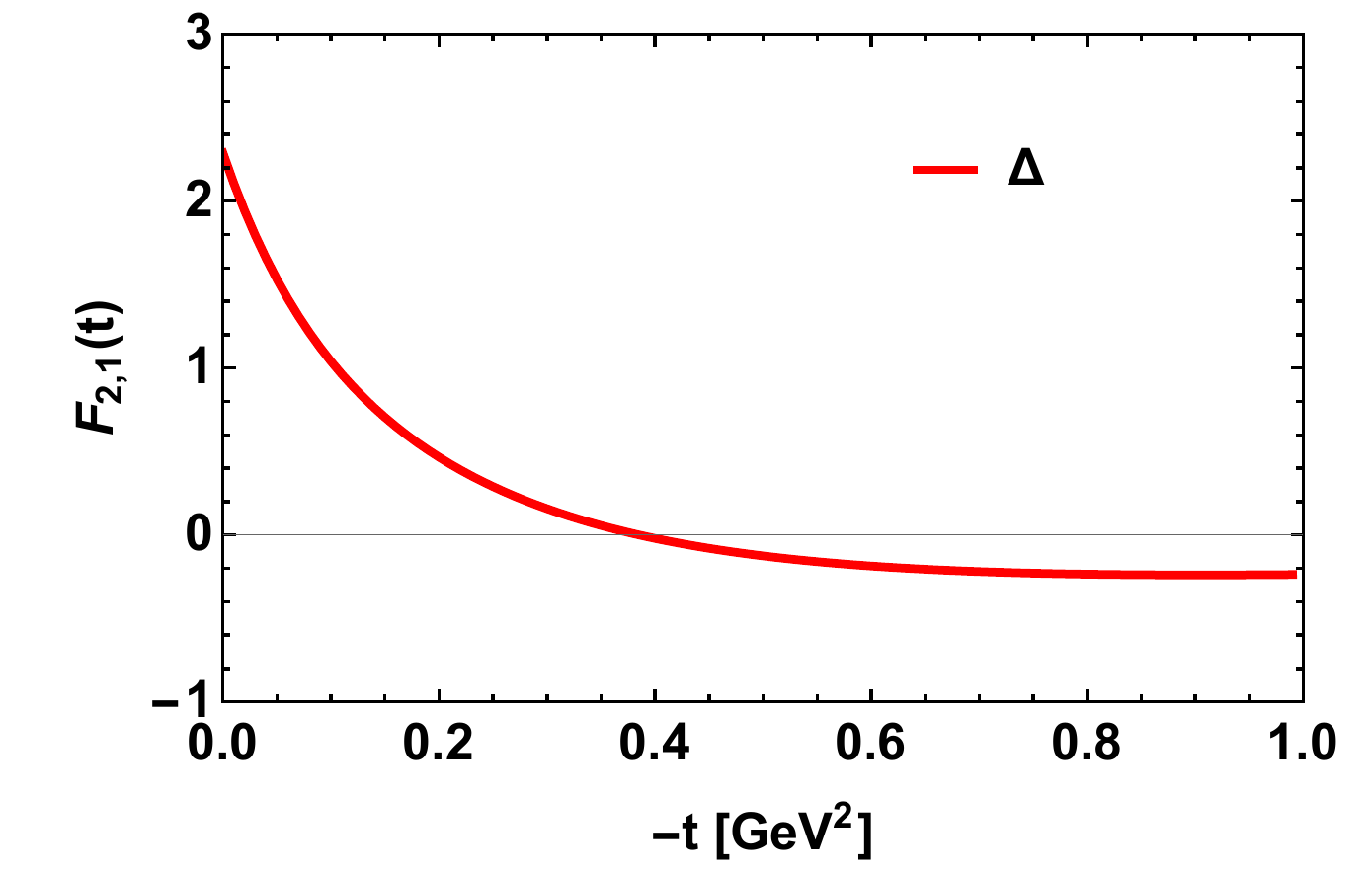}
\includegraphics[scale=0.629]{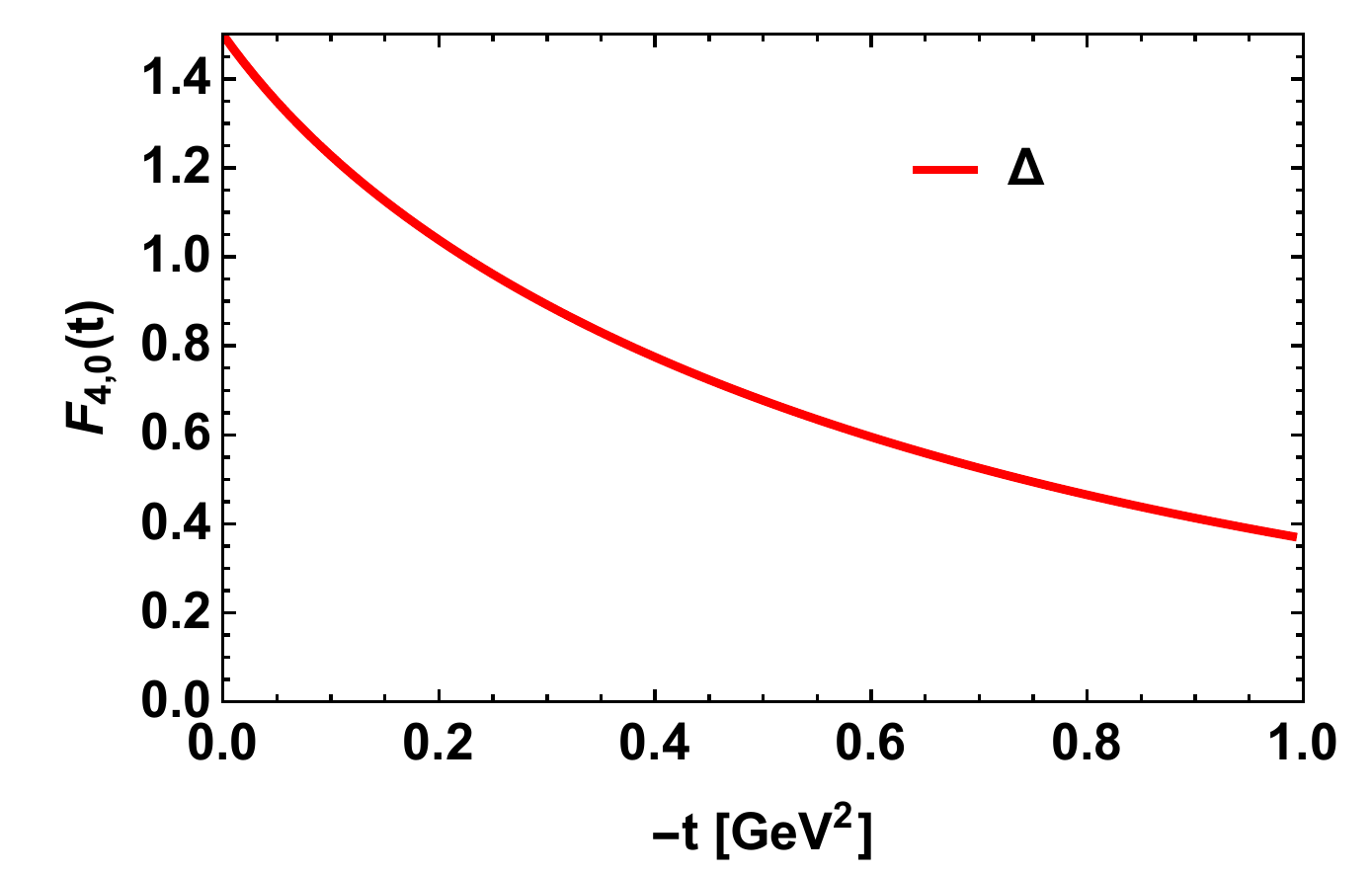}
\includegraphics[scale=0.629]{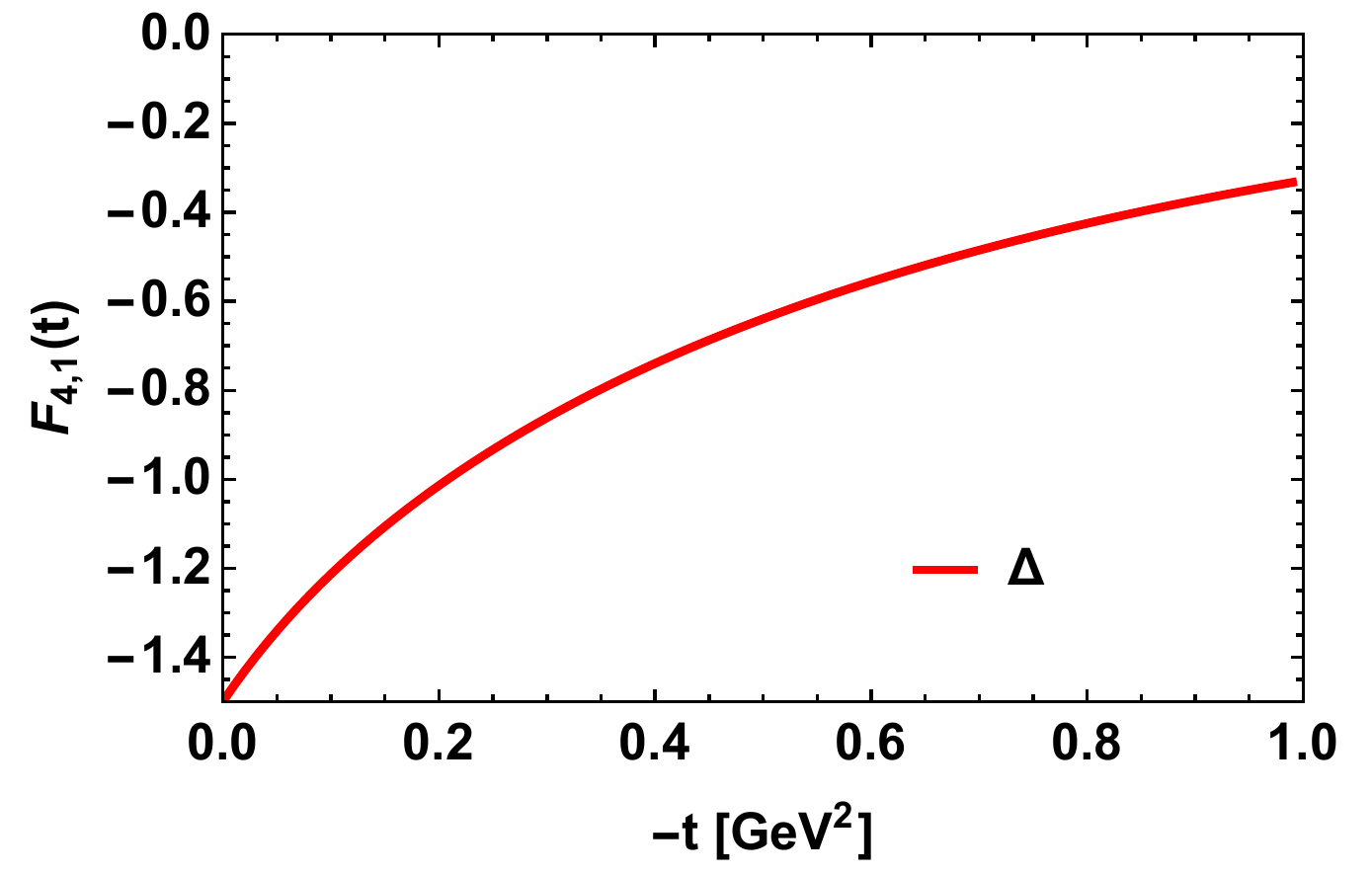}
\includegraphics[scale=0.629]{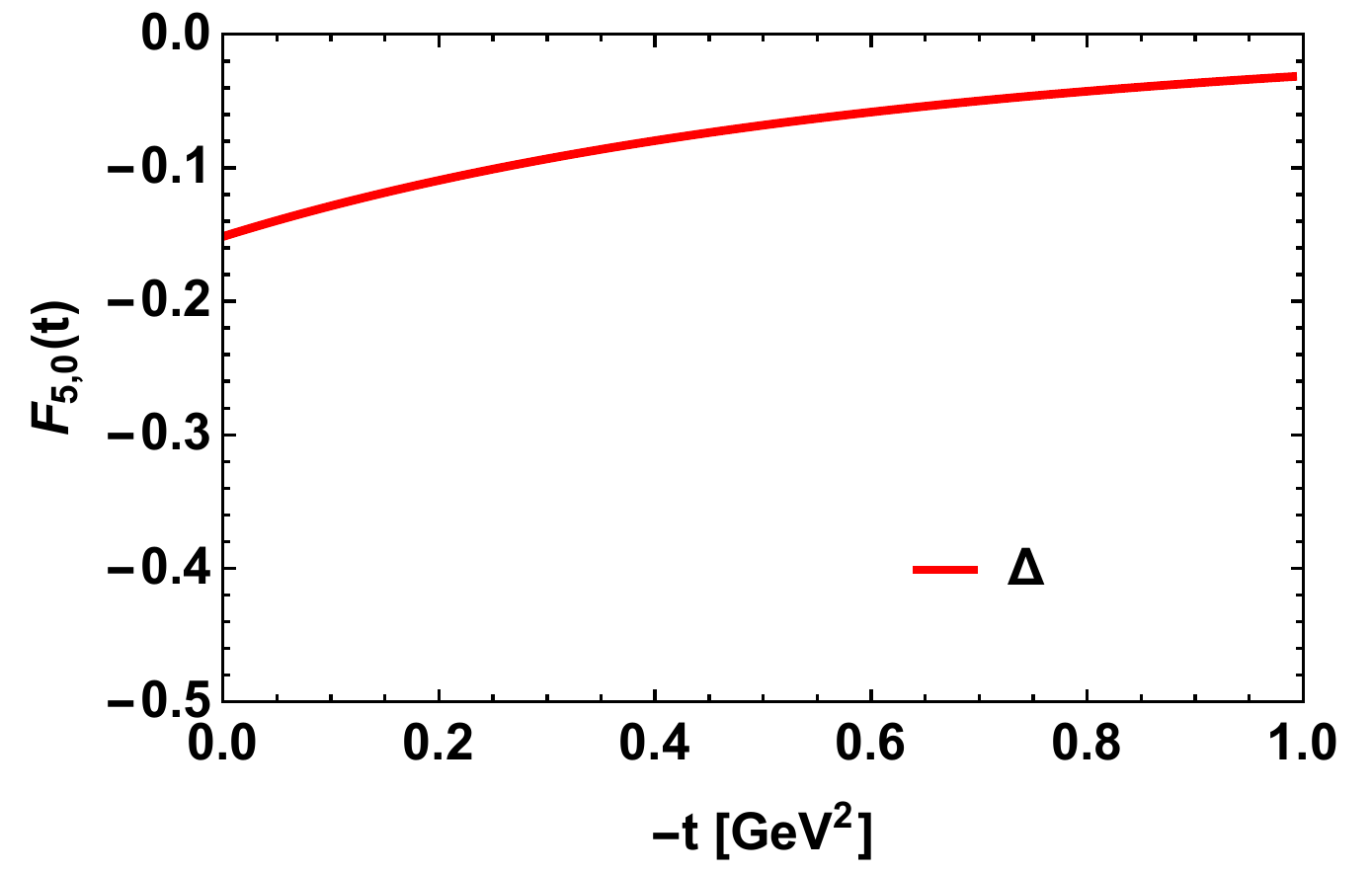}
\caption{Gravitational form factors of the $\Delta$ as a function of squared momentum transfer $t$.}
\label{fig:7}
\end{figure}

\section{Summary and conclusions \label{sec:5}}
In the present work, we aimed at providing the general form of the GFFs for a spin-3/2 baryon in terms of the multipole expansion and finding interesting relations between the EMT densities and these GFFs. We first defined the matrix elements of the EMT current given in Ref.~\cite{Cotogno:2019vjb} in terms of the GFFs, and then we expressed these GFFs by using the multipole expansion and related them with the EMT densities. The temporal component of the EMT density includes information on the mass $m$, its radius $\langle r^{2}_{E} \rangle$ and the gravitational quadrupole moment $\mathcal{Q}^{ij}_{\sigma'\sigma}$. The explicit relations between the multipole mass form factors $\mathcal{E}_{0,2}(t)$ and the multipole energy densities $\varepsilon_{0,2}(r)$ were provided. The spin density was related to the dipole and octupole spin form factors $\mathcal{J}_{1,3}(t)$. The integration of the spin density over $r$ yielded the constraint, which is analogous to a spin-1/2 baryon, for the dipole spin form factor $\mathcal{J}_{1}(t)$. The stress tensor, parametrized in terms of the pressure $p_{n}(r)$ and shear force $s_{n}(r)$ densities (for $n=0,2,3$), was related to the form factors $D_{n}(t)$. Interestingly, those densities should comply with the equilibrium equation~\eqref{eq:EMT_differential} to have the conserved EMT, which means that the respective pressures $p_{n}(r)$ have their own von Laue conditions and the corresponding generalized $D$-terms. Thus, we found a connection between the generalized $D$-terms $\mathcal{D}_{n}$ and the form factors $D_{n}(0)$. In addition, we obtained the expressions for the strong forces in terms of $p_{n}(r)$ and $s_{n}(r)$ inside a baryon and defined the generalized mechanical radii $\langle r^{2}_{n} \rangle_{\mathrm{mech}}$. 

To examine the general relations proposed in Sec.~\ref{sec:2}, we used the SU(2) Skyrme model based on the large-$N_{c}$ expansion. Since this model satisfies chiral symmetry and its spontaneous breaking, and described well the general properties of the nucleon GFFs, the model is suitable for investigating the GFFs of the $\Delta$. We first derived all the expressions for the EMT densities up to $\Omega^{2}\sim \mathcal{O}(N^{-2}_{c})$ and reproduced the large-$N_{c}$ relations proposed in Ref.~\cite{Panteleeva:2020ejw} within the Skyrme model. One of the remarkable relations was the null results of $p_{2}(r)$ and $s_{2}(r)$, which makes $\mathcal{D}_{2}$ vanishes. We also studied the $N_{c}$ behaviors of the GMFFs and the generalized $D$-terms.

In the Skyrme model, while the masses of the nucleon and the $\Delta$ were overestimated, they satisfied the constraint $\mathcal{E}_{0}(0)=F_{1,0}(0)=1$. We found that the energy density of the $\Delta$ is spreading more widely in comparison with that of the nucleon. We obtained the mass quadrupole moment for the $\Delta$ as $\mathcal{Q}^{ij}_{\sigma'\sigma} = -0.0181 Q^{ij}_{\sigma'\sigma} \mathrm{GeV}\cdot \mathrm{fm}^{2}$ that is related to the mass quadrupole form factor $\mathcal{E}_{2}(0)=0.34$. For the angular momentum density normalized by the corresponding baryon spin, we got the degenerate results on those for the nucleon and the $\Delta$. The angular momentum form factor for the $\Delta$ satisfied the constraint $\mathcal{J}_{1}(0)=F_{4,0}(0)/3=1/2$. Moreover, the null result of the octupole spin form factor $\mathcal{J}_{3}(t)=0$ gave the additional constraint $F_{4,1}(0)=-F_{4,0}(0)=-3/2$. When treating the pressure and shear force densities with included rotational corrections, we reconstructed the pressure densities $p_{n}(r)$ from the approximated shear force densities $s_{n}(r)$ through the equilibrium equation to comply with the von Laue condition. Utilizing this strategy, we understood that the $\Delta$ has  mechanically wide-spreading structure compared with the nucleon and obtained the generalized $D$-terms for the $\Delta$ as $\mathcal{D}_{0}=-3.53$, $\mathcal{D}_{2}=0$ and $\mathcal{D}_{3}=-0.50$. We predicted $t$ dependence on the GFFs and the GMFFs for the $\Delta$ in the range of $0<(-t)<1~\mathrm{GeV}^{2}$. We found that the typical sizes of the quadrupole form factors $\mathcal{E}_{2}(t), D_{2}(t)$ and $D_{3}(t)$ are relatively small in comparison with those of the monopole form factors $\mathcal{E}_{0}(t)$ and $ D_{0}(t)$.

As suggested in Ref.~\cite{Panteleeva:2020ejw} the lattice measurements of the $\Delta$ GFFs can be used to check whether $\Delta$-baryon is a rotating soliton. Here, we provided the first numerical estimates of corresponding GFFs in the soliton picture using the Skyrme model. We expect that the results of the lattice QCD or the theoretical model will soon come out.

\begin{acknowledgments}
The authors want to express M. V. Polyakov, H.-Ch. Kim and R.-H. Fang for valuable discussions.  J.-Y. Kim is supported by DAAD doctoral scholarship. B.-D. Sun is supported by the National Natural Sciences Foundations of China under the grant No.11947228 and China Postdoctoral Science Foundation under Grant No. 2019M662316. This work is also supported by the BMBF (Grant No. 05P18PCFP1).
\end{acknowledgments}

\appendix
\section{Breit frame formulae \label{Appendix:A}}
In the Breit frame the initial momentum $p$ and final momentum $p'$ have the relations $P^{\mu}=(p^{\mu}+p'^{\mu})/2=(E,0,0,0)$ and $\Delta^{\mu}= p'^{\mu}-p^{\mu}=(0,\bm{\Delta})$. The momentum squared is defined as $\Delta^{2}=-\bm{\Delta}^{2} = t = 4(m^{2}-E^{2})$.
Definition of the Rarita-Schwinger spinor is given by
\begin{align}
u^{\mu}=\sum_{}C^{\frac32\sigma}_{1\lambda\frac12s}u_s(p)\epsilon_\lambda^{\mu},
 \ \ \ \ \ \mathrm{with}  \ \ \ \ \
u_{s}(p)= \sqrt{{m+E}} 
\begin{pmatrix}
\phi_{s} \\
\frac{\bm{\sigma}\cdot\bm{p}}{m+E}\phi_{s}
\end{pmatrix},
\end{align}
with the two component Dirac spinor $\phi_{s}$.
The spin-1 vector $\epsilon^{\mu}_{\lambda}$ is defined by
\begin{align}
\epsilon_{\lambda}^{\mu}(p) = \left( -\frac{\bm{\Delta}\cdot \hat{\bm{\epsilon}}_{\lambda}}{2 m},\hat{\bm{\epsilon}}_{\lambda} + \frac{\bm{\Delta}\cdot \hat{\bm{\epsilon}}_{\lambda}}{4 m(m+E)}\bm{\Delta} \right), \ \ \ \epsilon_{\lambda'}^{\mu}(p') = \left( \frac{\bm{\Delta}\cdot \hat{\bm{\epsilon}}_{\lambda'}}{2 m},\hat{\bm{\epsilon}}_{\lambda'} + \frac{\bm{\Delta}\cdot \hat{\bm{\epsilon}}_{\lambda'}}{4 m(m+E)}\bm{\Delta} \right),
\end{align}
with 
\begin{align}
\hat{\bm{\epsilon}}_{+1}=\sqrt{\frac12}(-1,-i,0), \ \ \ \hat{\bm{\epsilon}}_{0}=\sqrt{\frac12}(0,0,1), \ \ \  \hat{\bm{\epsilon}}_{-1}=\sqrt{\frac12}(1,-i,0).
\end{align}
There are useful relations for Dirac field:
\begin{align}
\overline{u}_{s'}(p') \gamma^{0} u_{s}(p) &=2m \delta_{s's}, \cr
\overline{u}_{s'}(p') u_{s}(p) &= 2E\delta_{s's}, \cr
\overline{u}_{s'}(p') \frac{i}{2} (P^{0} \sigma^{0 \rho} \Delta_{\rho} + P^{0} \sigma^{0 \rho} \Delta_{\rho}) u_{s}(p) &= \Delta^{2}E\delta_{s's}, \cr
\overline{u}_{s'}(p') \frac{i}{2} (P^{i} \sigma^{0 \rho} \Delta_{\rho} + P^{0} \sigma^{i \rho} \Delta_{\rho}) u_{s}(p) &= 2i m E \epsilon^{ijk}\hat{S}^{(\frac{1}{2})j}_{s's}\Delta^{k}, \cr
\overline{u}_{s'}(p') \frac{i}{2} (P^{i} \sigma^{j \rho} \Delta_{\rho} + P^{j} \sigma^{i \rho} \Delta_{\rho}) u_{s}(p) &=0.
\end{align}
For vector field:
\begin{align}
(\hat{\bm{\epsilon}}^{*}_{\lambda'}\cdot \bm{\Delta})(\hat{\bm{\epsilon}}_{\lambda}\cdot \bm{\Delta}) &= -\frac{t}{3}\delta_{\lambda'\lambda}-\hat{Q}^{(1)kl}_{\lambda'\lambda}\Delta^{k}\Delta^{l}, \cr
\epsilon^{*}_{\lambda'} \cdot \epsilon_{\lambda} &= \left(\frac{t}{6m^{2}}-1\right)\delta_{\lambda'\lambda} + \frac{1}{2m^{2}} \hat{Q}^{(1)kl}_{\lambda'\lambda}\Delta^{k}\Delta^{l}, \cr
\epsilon^{0*}_{\lambda'} \epsilon^{0}_{\lambda} &= \frac{t}{12m^{2}}\delta_{\lambda'\lambda} + \frac{1}{4m^{2}} \hat{Q}^{(1)kl}_{\lambda'\lambda}\Delta^{k}\Delta^{l}, \cr
\epsilon^{*}_{\lambda'}\cdot {\Delta} &= -\frac{E}{m}\hat{\bm{\epsilon}}^{*}_{\lambda'} \cdot \bm{\Delta}, \cr
\epsilon_{\lambda}\cdot {\Delta} &= -\frac{E}{m}\hat{\bm{\epsilon}}_{\lambda} \cdot \bm{\Delta}, \cr
(\epsilon^{*}_{\lambda'}\cdot \Delta)(\epsilon_{\lambda}\cdot \Delta) &= -\frac{E^{2}}{m^{2}}\left(\frac{t}{3}\delta_{\lambda'\lambda}+\hat{Q}^{(1)kl}_{\lambda'\lambda}\Delta^{k}\Delta^{l}\right), \cr
{\epsilon}^{0*}_{\lambda'}({\epsilon}_{\lambda}\cdot {\Delta})+{\epsilon}^{0}_{\lambda}({\epsilon}^{*}_{\lambda'}\cdot {\Delta}) &= 0, \cr
\epsilon^{0*}_{\lambda'}\epsilon^{i}_{\lambda}+\epsilon^{i*}_{\lambda'}\epsilon^{0}_{\lambda}&=\frac{i}{2m} \epsilon^{kil}\Delta^{k}\hat{S}^{(1)l}_{\lambda'\lambda},\cr
\epsilon^{j*}_{\lambda'}\Delta^{i}(\epsilon_{\lambda}\cdot \Delta) + \epsilon^{j}_{\lambda}\Delta^{i}(\epsilon^{*}_{\lambda'}\cdot \Delta) &= -\frac{2E^{2}}{3M^{2}}\Delta^{i}\Delta^{j} \delta_{\lambda'\lambda} + \frac{2E}{m}\hat{Q}^{(1)jk}_{\lambda'\lambda}\Delta^{k}\Delta^{i} + \frac{E}{2m^{2}(m+E)}\hat{Q}^{(1)kl}_{\lambda'\lambda} \Delta^{k}\Delta^{l}\Delta^{i}\Delta^{j}, \cr
\epsilon^{i*}_{\lambda'}\epsilon^{j}_{\lambda}+\epsilon^{j*}_{\lambda'}\epsilon^{i}_{\lambda} &= \left(\frac{2}{3}\delta^{ij}+\frac{\Delta^{i}\Delta^{j}}{6m^{2}}\right)\delta_{\lambda'\lambda}-2\hat{Q}^{(1)ij}_{\lambda'\lambda}- \frac{1}{2m(m+E)}\left(\Delta^{i}\Delta^{k}\hat{Q}_{\lambda'\lambda}^{(1)kj}+\Delta^{j}\Delta^{k}\hat{Q}^{(1)ki}_{\lambda'\lambda}\right) \cr
&-\frac{1}{8m^{2}(m+E)^{2}}\Delta^{i}\Delta^{j}\Delta^{k}\Delta^{l}\hat{Q}^{(1)kl}_{\lambda'\lambda}.
\end{align}
The dipole- and quadrupole-spin operators for the spin-3/2 field are defined by
\begin{align}
\hat{S}^{i}_{\sigma'\sigma} &= \sum_{}C^{\frac32\sigma'}_{1\lambda'\frac12s'}C^{\frac32\sigma}_{1\lambda\frac12s} \left(-i\epsilon^{ijk}\hat{\epsilon}^{j*}_{\lambda'}\hat{\epsilon}^{k}_{\lambda}\delta_{s's} + \frac{\sigma_{s's}^{i}}{2}\delta_{\lambda'\lambda}\right),\cr
\hat{Q}^{ij}_{\sigma'\sigma} &= \sum_{}C^{\frac32\sigma'}_{1\lambda'\frac12s'}C^{\frac32\sigma}_{1\lambda\frac12s} \left[-\frac{1}{2}\left(\hat{\epsilon}^{i*}_{\lambda'}\hat{\epsilon}^{j}_{\lambda} + \hat{\epsilon}^{j*}_{\lambda'}\hat{\epsilon}^{i}_{\lambda}\right) -i\epsilon^{ilm}\hat{\epsilon}^{l*}_{\lambda'}\hat{\epsilon}^{m}_{\lambda} \frac{\sigma^{j}_{s's}}{2} -i\epsilon^{jlm}\hat{\epsilon}^{l*}_{\lambda'}\hat{\epsilon}^{m}_{\lambda} \frac{\sigma^{i}_{s's}}{2} \right].
\end{align}


\end{document}